\documentclass[12pt]{article}
\usepackage{graphicx}
\usepackage{subfigure}

\newcommand{\unit}[1]{\,{\mathrm{#1}}}
\newcommand{\ptt}{p_{\rm {t}}^{\rm {D}^{*+}}}

\newcommand{\ett}{\eta^{\rm {D}^{*+}}}

\newcommand{\dsdptt}{$\mathrm{d}\sigma/\mathrm{d}p_{\mathrm{t}}^{\mathrm{D}^{\ast +}}$}

\newcommand{\dsdetat}{$\mathrm{d}\sigma / \mathrm{d}|\eta^{\mathrm{D}^{\ast +}}|$}
\newcommand{\epem}  {\mathrm{e}^+\mathrm{e}^-}
\newcommand{\qqbar} {\mathrm{q}\bar{\mathrm{q}}}

\newcommand{\ccbar} {\mathrm{c}\bar{\mathrm{c}}}
\newcommand{\bbbar} {\mathrm{b}\bar{\mathrm{b}}}

\DeclareGraphicsExtensions{.eps,.ps,.eps.gz,.ps.gz}
\begin{document}

\begin{titlepage}

\begin{center}
EUROPEAN ORGANIZATION FOR NUCLEAR RESEARCH (CERN)
\end{center}

\begin{flushright}
\bigskip
\bigskip
\bigskip
CERN-EP/2003-02\\
10th January 2003 \\
\end{flushright}
\hfill\\
\bigskip
\bigskip
\bigskip
\begin{center}
{\LARGE \bf Measurement of the inclusive D$^{*\pm}$ production\\[8pt]
  in \boldmath $\gamma\gamma$ collisions at LEP}\\ \hfill\\
\bigskip
\bigskip
\bigskip 
{\large The ALEPH Collaboration$^{*)}$}
\hfill\\
\hfill\\
\end{center}
\bigskip
\bigskip
\bigskip

\begin{abstract}
\noindent
The inclusive production of D$^{*\pm}$ mesons in two-photon collisions
is measured with the ALEPH detector at $\epem$ centre-of-mass energies 
from 183$\unit{GeV}$ to 209$\unit{GeV}$. A total of $360 \pm 27$ 
D$^{*\pm}$ meson events were observed from an integrated luminosity of
$699\unit{pb^{-1}}$. Contributions from direct and single-resolved
processes are separated using the ratio of the transverse momentum 
$p_{\rm t}^{\rm D^{*\pm}}$ of the D$^{*\pm}$ to the visible invariant mass 
$W_{\mathrm{vis}}$ of the event.
Differential cross sections of D$^{*\pm}$ production as functions of 
 $p_{\rm t}^{\rm D^{*\pm}}$ and the pseudorapidity 
$|\eta^{\rm D^{*\pm}}| $ are measured in the range 
$ 2\unit{GeV}/c < p_{\rm t}^{\rm D^{*\pm}} < 12\unit{GeV}/c $ 
and $ |\eta^{\rm D^{*\pm}}| < 1.5 $. They are compared to next-to-leading
 order (NLO) perturbative QCD calculations.
The extrapolation of the integrated visible D$^{*\pm}$ cross section to the
total charm cross section, based on the Pythia Monte Carlo program, 
yields $ \sigma ( \epem \rightarrow \epem
\ccbar ) _ { <\sqrt{s}>=197\unit{GeV} } = 731 \pm 74_{\mathrm{stat}} \pm
47_{\mathrm{syst}} \pm 157_{\mathrm{extr}} \unit{pb} $.
\begin{center}

\bigskip
{\it Submitted to European Physical Journal C}
\end{center}

\bigskip

\end{abstract}
\vfill

~~~~~$^*)$ See next pages for the list of authors
\end{titlepage}
%\ifauthors
%------------------------------------------------------------------------
% authob12pt.tex
% authors' list for papers at LEP 1.5 and 2 energies
%-----------------------------------------------------------------------
\pagestyle{empty}
\newpage
\small
%
% remember the old settings
%
\newlength{\saveparskip}
\newlength{\savetextheight}
\newlength{\savetopmargin}
\newlength{\savetextwidth}
\newlength{\saveoddsidemargin}
\newlength{\savetopsep}
\setlength{\saveparskip}{\parskip}
\setlength{\savetextheight}{\textheight}
\setlength{\savetopmargin}{\topmargin}
\setlength{\savetextwidth}{\textwidth}
\setlength{\saveoddsidemargin}{\oddsidemargin}
\setlength{\savetopsep}{\topsep}
\typeout{\pretolerance=\the\pretolerance}
\edef\savepretolerance{\pretolerance}
%
% text dimensions for the author list
%
\setlength{\parskip}{0.0cm}
\setlength{\textheight}{25.0cm}
\setlength{\topmargin}{-1.5cm}
\setlength{\textwidth}{16 cm}
\setlength{\oddsidemargin}{-0.0cm}
\setlength{\topsep}{1mm}
\pretolerance=10000
\typeout{\pretolerance=\the\pretolerance}
%%\begin{document}
%\centerline{EUROPEAN ORGANIZATION FOR NUCLEAR RESEARCH}
%\centerline{EUROPEAN LABORATORY FOR PARTICLE PHYSICS (CERN)}
%\vspace{1cm}
%\begin{flushright}CERN-EP-2000-
%\9 October 2002 - last update
%\end{flushright}
\centerline{\large\bf The ALEPH Collaboration}
\footnotesize
\vspace{0.5cm}
{\raggedbottom
\begin{sloppypar}
\samepage\noindent
A.~Heister,
S.~Schael
\nopagebreak
\begin{center}
\parbox{15.5cm}{\sl\samepage
Physikalisches Institut das RWTH-Aachen, D-52056 Aachen, Germany}
\end{center}\end{sloppypar}
\vspace{2mm}
\begin{sloppypar}
\noindent
R.~Barate,
R.~Bruneli\`ere,
I.~De~Bonis,
D.~Decamp,
C.~Goy,
S.~Jezequel,
J.-P.~Lees,
F.~Martin,
E.~Merle,
\mbox{M.-N.~Minard},
B.~Pietrzyk,
B.~Trocm\'e
\nopagebreak
\begin{center}
\parbox{15.5cm}{\sl\samepage
Laboratoire de Physique des Particules (LAPP), IN$^{2}$P$^{3}$-CNRS,
F-74019 Annecy-le-Vieux Cedex, France}
\end{center}\end{sloppypar}
\vspace{2mm}
\begin{sloppypar}
\noindent
S.~Bravo,
M.P.~Casado,
M.~Chmeissani,
J.M.~Crespo,
E.~Fernandez,
M.~Fernandez-Bosman,
Ll.~Garrido,$^{15}$
M.~Martinez,
A.~Pacheco,
H.~Ruiz
\nopagebreak
\begin{center}
\parbox{15.5cm}{\sl\samepage
Institut de F\'{i}sica d'Altes Energies, Universitat Aut\`{o}noma
de Barcelona, E-08193 Bellaterra (Barcelona), Spain$^{7}$}
\end{center}\end{sloppypar}
\vspace{2mm}
\begin{sloppypar}
\noindent
A.~Colaleo,
D.~Creanza,
N.~De~Filippis,
M.~de~Palma,
G.~Iaselli,
G.~Maggi,
M.~Maggi,
S.~Nuzzo,
A.~Ranieri,
G.~Raso,$^{24}$
F.~Ruggieri,
G.~Selvaggi,
L.~Silvestris,
P.~Tempesta,
A.~Tricomi,$^{3}$
G.~Zito
\nopagebreak
\begin{center}
\parbox{15.5cm}{\sl\samepage
Dipartimento di Fisica, INFN Sezione di Bari, I-70126 Bari, Italy}
\end{center}\end{sloppypar}
\vspace{2mm}
\begin{sloppypar}
\noindent
X.~Huang,
J.~Lin,
Q. Ouyang,
T.~Wang,
Y.~Xie,
R.~Xu,
S.~Xue,
J.~Zhang,
L.~Zhang,
W.~Zhao
\nopagebreak
\begin{center}
\parbox{15.5cm}{\sl\samepage
Institute of High Energy Physics, Academia Sinica, Beijing, The People's
Republic of China$^{8}$}
\end{center}\end{sloppypar}
\vspace{2mm}
\begin{sloppypar}
\noindent
D.~Abbaneo,
P.~Azzurri,
T.~Barklow,$^{26}$
O.~Buchm\"uller,$^{26}$
M.~Cattaneo,
F.~Cerutti,
B.~Clerbaux,$^{23}$
H.~Drevermann,
R.W.~Forty,
M.~Frank,
F.~Gianotti,
J.B.~Hansen,
J.~Harvey,
D.E.~Hutchcroft,
P.~Janot,
B.~Jost,
M.~Kado,$^{2}$
P.~Mato,
A.~Moutoussi,
F.~Ranjard,
L.~Rolandi,
D.~Schlatter,
G.~Sguazzoni,
W.~Tejessy,
F.~Teubert,
A.~Valassi,
I.~Videau,
J.J.~Ward
\nopagebreak
\begin{center}
\parbox{15.5cm}{\sl\samepage
European Laboratory for Particle Physics (CERN), CH-1211 Geneva 23,
Switzerland}
\end{center}\end{sloppypar}
\vspace{2mm}
\begin{sloppypar}
\noindent
F.~Badaud,
S.~Dessagne,
A.~Falvard,$^{20}$
D.~Fayolle,
P.~Gay,
J.~Jousset,
B.~Michel,
S.~Monteil,
D.~Pallin,
J.M.~Pascolo,
P.~Perret
\nopagebreak
\begin{center}
\parbox{15.5cm}{\sl\samepage
Laboratoire de Physique Corpusculaire, Universit\'e Blaise Pascal,
IN$^{2}$P$^{3}$-CNRS, Clermont-Ferrand, F-63177 Aubi\`{e}re, France}
\end{center}\end{sloppypar}
\vspace{2mm}
\begin{sloppypar}
\noindent
J.D.~Hansen,
J.R.~Hansen,
P.H.~Hansen,
A.~Kraan,
B.S.~Nilsson
\nopagebreak
\begin{center}
\parbox{15.5cm}{\sl\samepage
Niels Bohr Institute, 2100 Copenhagen, DK-Denmark$^{9}$}
\end{center}\end{sloppypar}
\vspace{2mm}
\begin{sloppypar}
\noindent
A.~Kyriakis,
C.~Markou,
E.~Simopoulou,
A.~Vayaki,
K.~Zachariadou
\nopagebreak
\begin{center}
\parbox{15.5cm}{\sl\samepage
Nuclear Research Center Demokritos (NRCD), GR-15310 Attiki, Greece}
\end{center}\end{sloppypar}
\vspace{2mm}
\begin{sloppypar}
\noindent
A.~Blondel,$^{12}$
\mbox{J.-C.~Brient},
F.~Machefert,
A.~Roug\'{e},
M.~Swynghedauw,
R.~Tanaka
\linebreak
H.~Videau
\nopagebreak
\begin{center}
\parbox{15.5cm}{\sl\samepage
Laoratoire Leprince-Ringuet, Ecole
Polytechnique, IN$^{2}$P$^{3}$-CNRS, \mbox{F-91128} Palaiseau Cedex, France}
\end{center}\end{sloppypar}
\vspace{2mm}
\begin{sloppypar}
\noindent
V.~Ciulli,
E.~Focardi,
G.~Parrini
\nopagebreak
\begin{center}
\parbox{15.5cm}{\sl\samepage
Dipartimento di Fisica, Universit\`a di Firenze, INFN Sezione di Firenze,
I-50125 Firenze, Italy}
\end{center}\end{sloppypar}
\vspace{2mm}
\begin{sloppypar}
\noindent
A.~Antonelli,
M.~Antonelli,
G.~Bencivenni,
F.~Bossi,
G.~Capon,
V.~Chiarella,
P.~Laurelli,
G.~Mannocchi,$^{5}$
G.P.~Murtas,
L.~Passalacqua
\nopagebreak
\begin{center}
\parbox{15.5cm}{\sl\samepage
Laboratori Nazionali dell'INFN (LNF-INFN), I-00044 Frascati, Italy}
\end{center}\end{sloppypar}
\vspace{2mm}
%\pagebreak
\begin{sloppypar}
\noindent
J.~Kennedy,
J.G.~Lynch,
P.~Negus,
V.~O'Shea,
A.S.~Thompson
\nopagebreak
\begin{center}
\parbox{15.5cm}{\sl\samepage
Department of Physics and Astronomy, University of Glasgow, Glasgow G12
8QQ,United Kingdom$^{10}$}
\end{center}\end{sloppypar}
\vspace{2mm}
%\pagebreak
\begin{sloppypar}
\noindent
S.~Wasserbaech
\nopagebreak
\begin{center}
\parbox{15.5cm}{\sl\samepage
Department of Physics, Haverford College, Haverford, PA 19041-1392, U.S.A.}
\end{center}\end{sloppypar}
\vspace{2mm}
%\pagebreak
\begin{sloppypar}
\noindent
R.~Cavanaugh,$^{4}$
S.~Dhamotharan,$^{21}$
C.~Geweniger,
P.~Hanke,
V.~Hepp,
E.E.~Kluge,
G.~Leibenguth,
A.~Putzer,
H.~Stenzel,
K.~Tittel,
M.~Wunsch$^{19}$
\nopagebreak
\begin{center}
\parbox{15.5cm}{\sl\samepage
Kirchhoff-Institut f\"ur Physik, Universit\"at Heidelberg, D-69120
Heidelberg, Germany$^{16}$}
\end{center}\end{sloppypar}
\vspace{2mm}
\begin{sloppypar}
\noindent
R.~Beuselinck,
W.~Cameron,
G.~Davies,
P.J.~Dornan,
M.~Girone,$^{1}$
R.D.~Hill,
N.~Marinelli,
J.~Nowell,
S.A.~Rutherford,
J.K.~Sedgbeer,
J.C.~Thompson,$^{14}$
R.~White
\nopagebreak
\begin{center}
\parbox{15.5cm}{\sl\samepage
Department of Physics, Imperial College, London SW7 2BZ,
United Kingdom$^{10}$}
\end{center}\end{sloppypar}
\vspace{2mm}
\begin{sloppypar}
\noindent
V.M.~Ghete,
P.~Girtler,
E.~Kneringer,
D.~Kuhn,
G.~Rudolph
\nopagebreak
\begin{center}
\parbox{15.5cm}{\sl\samepage
Institut f\"ur Experimentalphysik, Universit\"at Innsbruck, A-6020
Innsbruck, Austria$^{18}$}
\end{center}\end{sloppypar}
\vspace{2mm}
\begin{sloppypar}
\noindent
E.~Bouhova-Thacker,
C.K.~Bowdery,
D.P.~Clarke,
G.~Ellis,
A.J.~Finch,
F.~Foster,
G.~Hughes,
R.W.L.~Jones,
M.R.~Pearson,
N.A.~Robertson,
M.~Smizanska
\nopagebreak
\begin{center}
\parbox{15.5cm}{\sl\samepage
Department of Physics, University of Lancaster, Lancaster LA1 4YB,
United Kingdom$^{10}$}
\end{center}\end{sloppypar}
\vspace{2mm}
\begin{sloppypar}
\noindent
O.~van~der~Aa,
C.~Delaere,$^{28}$
V.~Lemaitre$^{29}$
\nopagebreak
\begin{center}
\parbox{15.5cm}{\sl\samepage
Institut de Physique Nucl\'eaire, D\'epartement de Physique, Universit\'e Catholique de Louvain, 1348 Louvain-la-Neuve, Belgium}
\end{center}\end{sloppypar}
\vspace{2mm}
\begin{sloppypar}
\noindent
U.~Blumenschein,
F.~H\"olldorfer,
K.~Jakobs,
F.~Kayser,
K.~Kleinknecht,
A.-S.~M\"uller,
B.~Renk,
H.-G.~Sander,
S.~Schmeling,
H.~Wachsmuth,
C.~Zeitnitz,
T.~Ziegler
\nopagebreak
\begin{center}
\parbox{15.5cm}{\sl\samepage
Institut f\"ur Physik, Universit\"at Mainz, D-55099 Mainz, Germany$^{16}$}
\end{center}\end{sloppypar}
\vspace{2mm}
\begin{sloppypar}
\noindent
A.~Bonissent,
P.~Coyle,
C.~Curtil,
A.~Ealet,
D.~Fouchez,
P.~Payre,
A.~Tilquin
\nopagebreak
\begin{center}
\parbox{15.5cm}{\sl\samepage
Centre de Physique des Particules de Marseille, Univ M\'editerran\'ee,
IN$^{2}$P$^{3}$-CNRS, F-13288 Marseille, France}
\end{center}\end{sloppypar}
\vspace{2mm}
\begin{sloppypar}
\noindent
F.~Ragusa
\nopagebreak
\begin{center}
\parbox{15.5cm}{\sl\samepage
Dipartimento di Fisica, Universit\`a di Milano e INFN Sezione di
Milano, I-20133 Milano, Italy.}
\end{center}\end{sloppypar}
\vspace{2mm}
\begin{sloppypar}
\noindent
A.~David,
H.~Dietl,
G.~Ganis,$^{27}$
K.~H\"uttmann,
G.~L\"utjens,
W.~M\"anner,
\mbox{H.-G.~Moser},
R.~Settles,
G.~Wolf
\nopagebreak
\begin{center}
\parbox{15.5cm}{\sl\samepage
Max-Planck-Institut f\"ur Physik, Werner-Heisenberg-Institut,
D-80805 M\"unchen, Germany\footnotemark[16]}
\end{center}\end{sloppypar}
\vspace{2mm}
\begin{sloppypar}
\noindent
J.~Boucrot,
O.~Callot,
M.~Davier,
L.~Duflot,
\mbox{J.-F.~Grivaz},
Ph.~Heusse,
A.~Jacholkowska,$^{6}$
L.~Serin,
\mbox{J.-J.~Veillet},
C.~Yuan
\nopagebreak
\begin{center}
\parbox{15.5cm}{\sl\samepage
Laboratoire de l'Acc\'el\'erateur Lin\'eaire, Universit\'e de Paris-Sud,
IN$^{2}$P$^{3}$-CNRS, F-91898 Orsay Cedex, France}
\end{center}\end{sloppypar}
\vspace{2mm}
\begin{sloppypar}
\noindent
%\samepage
G.~Bagliesi,
T.~Boccali,
L.~Fo\`a,
A.~Giammanco,
A.~Giassi,
F.~Ligabue,
A.~Messineo,
F.~Palla,
G.~Sanguinetti,
A.~Sciab\`a,
R.~Tenchini,$^{1}$
A.~Venturi,$^{1}$
P.G.~Verdini
\samepage
\begin{center}
\parbox{15.5cm}{\sl\samepage
Dipartimento di Fisica dell'Universit\`a, INFN Sezione di Pisa,
e Scuola Normale Superiore, I-56010 Pisa, Italy}
\end{center}\end{sloppypar}
\vspace{2mm}
\begin{sloppypar}
\noindent
O.~Awunor,
G.A.~Blair,
G.~Cowan,
A.~Garcia-Bellido,
M.G.~Green,
L.T.~Jones,
T.~Medcalf,
A.~Misiejuk,
J.A.~Strong,
P.~Teixeira-Dias
\nopagebreak
\begin{center}
\parbox{15.5cm}{\sl\samepage
Department of Physics, Royal Holloway \& Bedford New College,
University of London, Egham, Surrey TW20 OEX, United Kingdom$^{10}$}
\end{center}\end{sloppypar}
\vspace{2mm}
\begin{sloppypar}
\noindent
R.W.~Clifft,
T.R.~Edgecock,
P.R.~Norton,
I.R.~Tomalin
\nopagebreak
\begin{center}
\parbox{15.5cm}{\sl\samepage
Particle Physics Dept., Rutherford Appleton Laboratory,
Chilton, Didcot, Oxon OX11 OQX, United Kingdom$^{10}$}
\end{center}\end{sloppypar}
\vspace{2mm}
%\pagebreak
\begin{sloppypar}
\noindent
\mbox{B.~Bloch-Devaux},
D.~Boumediene,
P.~Colas,
B.~Fabbro,
E.~Lan\c{c}on,
\mbox{M.-C.~Lemaire},
E.~Locci,
P.~Perez,
J.~Rander,
B.~Tuchming,
B.~Vallage
\nopagebreak
\begin{center}
\parbox{15.5cm}{\sl\samepage
CEA, DAPNIA/Service de Physique des Particules,
CE-Saclay, F-91191 Gif-sur-Yvette Cedex, France$^{17}$}
\end{center}\end{sloppypar}
%\nopagebreak
\vspace{2mm}
\begin{sloppypar}
\noindent
N.~Konstantinidis,
A.M.~Litke,
G.~Taylor
\nopagebreak
\begin{center}
\parbox{15.5cm}{\sl\samepage
Institute for Particle Physics, University of California at
Santa Cruz, Santa Cruz, CA 95064, USA$^{22}$}
\end{center}\end{sloppypar}
%\pagebreak
\vspace{2mm}
\begin{sloppypar}
\noindent
C.N.~Booth,
S.~Cartwright,
F.~Combley,$^{25}$
P.N.~Hodgson,
M.~Lehto,
L.F.~Thompson
\nopagebreak
\begin{center}
\parbox{15.5cm}{\sl\samepage
Department of Physics, University of Sheffield, Sheffield S3 7RH,
United Kingdom$^{10}$}
\end{center}\end{sloppypar}
\vspace{2mm}
\begin{sloppypar}
\noindent
A.~B\"ohrer,
S.~Brandt,
C.~Grupen,
J.~Hess,
A.~Ngac,
G.~Prange,
U.~Sieler$^{30}$
\nopagebreak
\begin{center}
\parbox{15.5cm}{\sl\samepage
Fachbereich Physik, Universit\"at Siegen, D-57068 Siegen, Germany$^{16}$}
\end{center}\end{sloppypar}
\vspace{2mm}
\begin{sloppypar}
\noindent
C.~Borean,
G.~Giannini
\nopagebreak
\begin{center}
\parbox{15.5cm}{\sl\samepage
Dipartimento di Fisica, Universit\`a di Trieste e INFN Sezione di Trieste,
I-34127 Trieste, Italy}
\end{center}\end{sloppypar}
\vspace{2mm}
\begin{sloppypar}
\noindent
H.~He,
J.~Putz,
J.~Rothberg
\nopagebreak
\begin{center}
\parbox{15.5cm}{\sl\samepage
Experimental Elementary Particle Physics, University of Washington, Seattle,
WA 98195 U.S.A.}
\end{center}\end{sloppypar}
\vspace{2mm}
\begin{sloppypar}
\noindent
S.R.~Armstrong,
K.~Berkelman,
K.~Cranmer,
D.P.S.~Ferguson,
Y.~Gao,$^{13}$
S.~Gonz\'{a}lez,
O.J.~Hayes,
H.~Hu,
S.~Jin,
J.~Kile,
P.A.~McNamara III,
J.~Nielsen,
Y.B.~Pan,
\mbox{J.H.~von~Wimmersperg-Toeller}, 
W.~Wiedenmann,
J.~Wu,
Sau~Lan~Wu,
X.~Wu,
G.~Zobernig
\nopagebreak
\begin{center}
\parbox{15.5cm}{\sl\samepage
Department of Physics, University of Wisconsin, Madison, WI 53706,
USA$^{11}$}
\end{center}\end{sloppypar}
\vspace{2mm}
\begin{sloppypar}
\noindent
G.~Dissertori
\nopagebreak
\begin{center}
\parbox{15.5cm}{\sl\samepage
Institute for Particle Physics, ETH H\"onggerberg, 8093 Z\"urich,
Switzerland.}
\end{center}\end{sloppypar}
}
\footnotetext[1]{Also at CERN, 1211 Geneva 23, Switzerland.}
\footnotetext[2]{Now at Fermilab, PO Box 500, MS 352, Batavia, IL 60510, USA}
\footnotetext[3]{Also at Dipartimento di Fisica di Catania and INFN Sezione di
 Catania, 95129 Catania, Italy.}
\footnotetext[4]{Now at University of Florida, Department of Physics, Gainesville, Florida 32611-8440, USA}
\footnotetext[5]{Also Istituto di Cosmo-Geofisica del C.N.R., Torino,
Italy.}
\footnotetext[6]{Also at Groupe d'Astroparticules de Montpellier, Universit\'{e} de Montpellier II, 34095, Montpellier, France.}
\footnotetext[7]{Supported by CICYT, Spain.}
\footnotetext[8]{Supported by the National Science Foundation of China.}
\footnotetext[9]{Supported by the Danish Natural Science Research Council.}
\footnotetext[10]{Supported by the UK Particle Physics and Astronomy Research
Council.}
\footnotetext[11]{Supported by the US Department of Energy, grant
DE-FG0295-ER40896.}
\footnotetext[12]{Now at Departement de Physique Corpusculaire, Universit\'e de
Gen\`eve, 1211 Gen\`eve 4, Switzerland.}
\footnotetext[13]{Also at Department of Physics, Tsinghua University, Beijing, The People's Republic of China.}
\footnotetext[14]{Supported by the Leverhulme Trust.}
\footnotetext[15]{Permanent address: Universitat de Barcelona, 08208 Barcelona,
Spain.}
\footnotetext[16]{Supported by Bundesministerium f\"ur Bildung
und Forschung, Germany.}
\footnotetext[17]{Supported by the Direction des Sciences de la
Mati\`ere, C.E.A.}
\footnotetext[18]{Supported by the Austrian Ministry for Science and Transport.}
\footnotetext[19]{Now at SAP AG, 69185 Walldorf, Germany}
\footnotetext[20]{Now at Groupe d' Astroparticules de Montpellier, Universit\'e de Montpellier II, 34095 Montpellier, France.}
\footnotetext[21]{Now at BNP Paribas, 60325 Frankfurt am Mainz, Germany}
\footnotetext[22]{Supported by the US Department of Energy,
grant DE-FG03-92ER40689.}
\footnotetext[23]{Now at Institut Inter-universitaire des hautes Energies (IIHE), CP 230, Universit\'{e} Libre de Bruxelles, 1050 Bruxelles, Belgique}
\footnotetext[24]{Also at Dipartimento di Fisica e Tecnologie Relative, Universit\`a di Palermo, Palermo, Italy.}
\footnotetext[25]{Deceased.}
\footnotetext[26]{Now at SLAC, Stanford, CA 94309, U.S.A}
\footnotetext[27]{Now at INFN Sezione di Roma II, Dipartimento di Fisica, Universit\`a di Roma Tor Vergata, 00133 Roma, Italy.}
\footnotetext[28]{Research Fellow of the Belgium FNRS}
\footnotetext[29]{Research Associate of the Belgium FNRS}
\footnotetext[30]{Now at Verdi Information Consult GmbH, 53757 Sankt Augustin, Germany}  
\setlength{\parskip}{\saveparskip}
\setlength{\textheight}{\savetextheight}
\setlength{\topmargin}{\savetopmargin}
\setlength{\textwidth}{\savetextwidth}
\setlength{\oddsidemargin}{\saveoddsidemargin}
\setlength{\topsep}{\savetopsep}
\pretolerance=100
\typeout{\pretolerance=\the\pretolerance}
\normalsize
\newpage
\pagestyle{plain}
\setcounter{page}{1}
%\fi
\newpage
\pagenumbering{arabic}
\setcounter{page}{1}
\section{Introduction}
\label{sec:intro}
\begin{figure}
  \begin{center} 
        \includegraphics[width=0.45\textwidth]{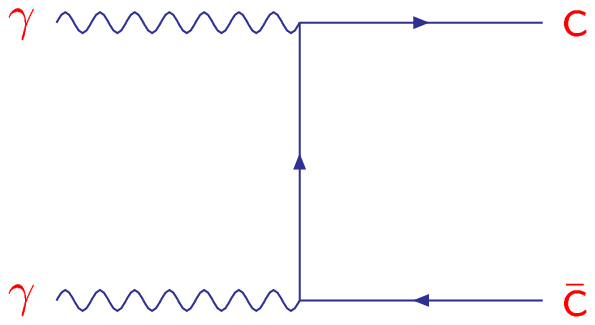}
       \hfill
        \includegraphics[width=0.45\textwidth]{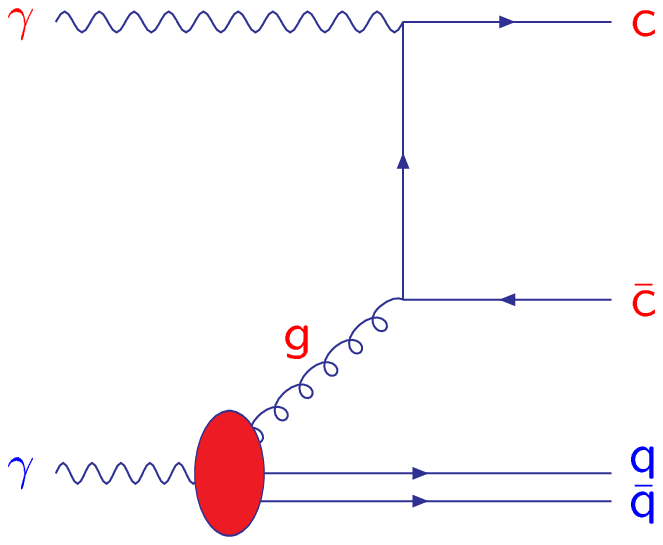}
   \end{center}

    {\hspace*{23mm} a) Direct process \hspace{45mm} b) Single-resolved process}

    \caption{Main contributions to charm production in $\gamma\gamma$ events.}
   \label{fig:main_proc}
\end{figure}

Heavy flavour production in two-photon events at LEP 2 centre-of-mass 
energies is dominated by charm production processes in which both of the 
photons couple directly (\emph{direct processes}) or in which one
photon couples directly and the other appears resolved
(\emph{single-resolved processes}) (Fig.~\ref{fig:main_proc}) \cite{zerwas:93}. 
These two contributions are of the same order of magnitude within the 
experimental acceptance.
 Because the single-resolved process is dominated by $\gamma g$ fusion,  
 the measurement of the cross section can give access to the gluon content of 
 the photon. Moreover, the large masses of the c
 and b  quarks provide a cutoff for perturbative QCD calculations,
 allowing a good test of QCD predictions for the corresponding reactions. 
Contributions from processes in which both photons appear resolved 
(\emph{double-resolved processes}) are suppressed by more than two orders 
of magnitude compared to the total cross section \cite{zerwas:93}. 
The production of b quark is expected to be suppressed by a large factor
compared to charm quark because of the heavier mass and smaller
absolute charge.

 In the present analysis charm production is measured in two
steps. A high-purity $\gamma\gamma$ sample is first selected, 
 then examined for its charm content via reconstruction of
D$^{*+}$ mesons in their decay to $\textrm{D}^0 \pi^+$.
  This letter is organized as follows. A short description of the ALEPH 
detector is given in Section~\ref{sec:aleph}. Monte Carlo simulations for 
signal and background processes are described in Section~\ref{sec:MC}. 
In Section~\ref{sec:select}, event selection and reconstruction of D$^{*+}$ 
mesons are discussed. The results of the analysis are presented in 
Section~\ref{sec:XS}. 
Finally, in Section~\ref{sec:conclusions} a summary is given.
Throughout this letter charge-conjugated particles and their decays are 
implicitly included.
\section{ALEPH Detector}
\label{sec:aleph}
 The ALEPH detector has been described in detail in
\cite{aleph1990,aleph1995}. Here, only the parts essential to the
present analysis are covered briefly. 
The central part of the ALEPH detector is 
dedicated to the reconstruction of the trajectories of charged
particles. The trajectory of a charged particle emerging from the 
interaction point is measured by a two-layer silicon strip 
vertex detector (VDET), a cylindrical drift chamber (ITC) and a large time 
projection chamber (TPC). The three tracking detectors are immersed in a 
$1.5\unit{T}$ axial magnetic field provided by a superconducting solenoidal 
coil. Together they measure charged particle transverse 
momenta with a resolution 
of $\delta p_{\mathrm t}/p_{\mathrm t} = 6 \times 10^{-4} p_{\mathrm t} 
\oplus 0.005$ ($p_{\mathrm t}$ in GeV/$c$). The TPC also provides a 
measurement of the specific ionization 
${\mathrm d}E/{\mathrm d}x_{\mathrm{meas}}$. An estimator $\chi_{\mathrm h} = 
({\mathrm d}E/{\mathrm d}x_{\mathrm{meas}}-
{\mathrm d}E/{\mathrm d}x_{\mathrm{exp,h}}) / \sigma_{\mathrm{exp,h}}$ 
is formed to test a particle hypothesis, where 
${\mathrm d}E/{\mathrm d}x_{\mathrm{exp,h}}$ and 
$\sigma_{\mathrm{exp,h}}$ denote the expected specific ionization and 
the estimated uncertainty for the particle hypothesis h, respectively.
A mass hypothesis may be tested by means of
the $\chi_{\mathrm h}$ values themselves or by calculating $\chi_{\mathrm h}^2$
confidence levels $P_{\mathrm h}$.
 
Photons are identified in the electromagnetic calorimeter 
(ECAL), situated between the TPC and the coil. The ECAL 
is a lead/proportional-tube 
sampling calorimeter segmented in $0.9^{\circ} \times 0.9^{\circ}$ projective 
towers and read out in three sections in depth. It has a total thickness 
of 22 radiation lengths and yields a relative energy resolution of 
$0.18/\sqrt{E} + 0.009$, with $E$ in GeV, for isolated photons. Electrons are 
identified by their transverse and longitudinal shower profiles in 
ECAL and their specific ionization in the TPC.

The iron return yoke is instrumented with 23 layers of streamer tubes and 
forms the hadron calorimeter (HCAL). The latter provides a relative 
energy resolution of charged and neutral hadrons of $0.85/\sqrt{E}$, 
with $E$ in GeV. Muons are distinguished from hadrons by their characteristic 
pattern in HCAL and by the muon chambers, composed of two double-layers 
of streamer tubes outside HCAL. 

 Two small-angle calorimeters, the luminosity calorimeter (LCAL) and
 the silicon luminosity calorimeter (SICAL),
are particularly important for this analysis to veto events with  detected 
scattered electrons. The LCAL is a lead/proportional-tube calorimeter, 
similar to ECAL, placed around the beam pipe at each end of the detector. 
It monitors angles from 45 to 160 mrad with an energy resolution of 
$0.15 \sqrt{E(\rm GeV)}$.
 The SICAL uses 12 silicon/tungsten layers to sample showers. It is 
mounted around the beam pipe in front of the LCAL, covering angles from
34 to 58 mrad, with an energy resolution of $0.225 \sqrt{E(\rm GeV)}$.

The information from the tracking detectors and the calorimeters are 
combined in an energy-flow algorithm~\cite{aleph1995}. For each event, the 
algorithm provides a set of charged and neutral reconstructed particles, 
called {\em energy-flow objects}.

\section{Monte Carlo Simulations}
\label{sec:MC}
In order to simulate the process 
$\epem \rightarrow \epem \gamma\gamma \rightarrow \epem \ccbar \rightarrow 
\epem \rm D^{*\pm} X $, the leading-order (LO) PYTHIA 6.121 Monte Carlo 
\cite{sjoestrand:94} is used. Events are generated at $\epem$ 
centre-of-mass energies ranging from 183$\unit{GeV}$ to 209$\unit{GeV}$ using the corresponding 
 integrated luminosities for weighting. Two different samples, direct and  
single-resolved processes, were generated for each of the considered  
D$^{*+}$ decay modes using matrix elements for the massive charm quark. 
The charm quark mass $m_{\rm c}$ is chosen to be 1.5$\unit{GeV/c^{2}}$
 and the parameter $\Lambda_{\rm QCD}$ is set to 0.291$\unit{GeV/c^{2}}$ . 
 The $\gamma\gamma$ invariant mass $W_{\gamma\gamma}$ is required to be at least 
 3.875$\unit{GeV/c^{2}}$, which is the D$\bar{\rm D}$ threshold. In order to ensure that both 
photons are quasi-real, the maximum  squared four-momentum transfer 
$Q^{2}_{\rm max}$ is limited to 4.5$\unit{GeV/c^{2}}$.
In the single-resolved process, the SaS-1D \cite{sas-1d} parametrization is used for 
the partonic distribution of the resolved photon. The Peterson et al.\ 
parametrization \cite{peterson} is adopted as the fragmentation function of the charm 
quark with the nonperturbative parameter $\epsilon_{\rm c} = 0.031$. 
The background process $\epem \rightarrow \epem \gamma\gamma \rightarrow 
\epem \bbbar $ is simulated using PYTHIA 6.121
with $W_{\gamma\gamma}$ being required to be at least 10.5$\unit{GeV/c^{2}}$,
 which is the B$\bar{\rm B}$ threshold. The b quark mass is set to  
4.5$\unit{GeV/c^{2}}$. Again the Peterson et al.\ parametrization  
is adopted with $\epsilon_{\rm b} = 0.0035$. 
Other possible background processes have been simulated using appropriate
Monte Carlo generators as listed in Table~\ref{tab:backgr_MC}.

\section{Event Selection and Reconstruction of D$^{*+}$ Mesons}
\label{sec:select}

\subsection{Selection of $\gamma\gamma$ Events}
\label{sec:selection}
The data analyzed were collected by the ALEPH detector
at $\epem$ centre-of-mass energies ranging from 183$\unit{GeV}$ to 209$\unit{GeV}$
 with an integrated luminosity $ \mathcal{L} =
699\unit{pb^{-1}}$. The event variables used for the event preselection are
based on the ALEPH energy-flow objects.
The following cuts, derived from Monte Carlo studies, were applied to select 
two-photon events.
\begin{itemize}
\itemsep -4pt
\item The event must contain at least 3 charged particles. This cut reduces 
      the background from leptonic events.
\item The visible invariant mass $W_{\mathrm{vis}}$ of the event 
must lie between 4$\unit{GeV}/c^{2}$ and 55$\unit{GeV}/c^{2}$ while the 
total energy of charged particles
$E_{\mathrm{ch}}$ should not exceed 35$\unit{GeV}$ in order to reject the
$\epem$ annihilation background. 
\item The visible transverse momentum $p_{\rm t,\mathrm{vis}}$ of the event 
is required to be less than 8$\unit{GeV}/c$, as the $p_{\rm t,\mathrm{vis}}$ 
distribution
has a much longer tail for all considered background processes.
\item To reject further background processes a cut combining the 
number of charged tracks and the visible energy $E_{\mathrm{vis}}$ of the event
 is applied:  $N_{\mathrm{ch}} < 40 - \frac{2}{3} E_{\mathrm{vis}} (\rm{GeV})$.
\item Finally, in order to retain only events with almost on-shell photons 
an anti-tagging condition was applied, i.e., tagged events were rejected.  
A tag in this analysis is defined as an energy-flow object in the luminosity 
calorimeters (LCAL and SICAL) with an energy of at least 30$\unit{GeV}$.
\end{itemize}
  This selection retains a sample of 4.9 million events. Monte
Carlo studies of possible background sources predict a $\gamma\gamma$
purity of $98.8 \% $.

\subsection{Reconstruction of D$^{*+}$ Mesons}
\label{sec:dstar}
Charm quarks are detected using exclusively reconstructed 
D$^{*+}$ mesons which decay via D$^{*+} \rightarrow \textrm{D}^0 \pi^+$, with
the D$^0$ being identified in three decay modes, (1) $\rm K^{-} \pi^+$,
(2) $\rm K^{-} \pi^+ \pi^0$, and (3) $\rm K^{-} \pi^+ \pi^- \pi^+$.  
As a basis for possible K$^{\pm}$ and $\pi^{\pm}$ candidates
reconstructed tracks of charged particles which fulfill the following
quality conditions are used:
\begin{tabbing}
  \hspace*{10pt} \= \hspace{10mm} \= \hspace{35mm} \= rest  \kill 
  \> $p$ \> $ > 100\unit{MeV}/c$ \> (momentum of track),\\
  \> $|d_0|$ \> $ < 2\unit{cm}$ \> (distance to beam axis at closest approach),\\
  \> $|z_0|$ \> $ < 8\unit{cm}$ \> ($z$ coordinate at closest approach),\\
  \> $N_{\mathrm{TPC}}$ \> $ \geq 4$ \> (number of hits in TPC),\\
  \> $|{\cos \theta}|$ \> $ < 0.94$ \> ($\theta =$ polar angle with respect to beam axis).
\end{tabbing}
A track surviving these cuts is classified as a kaon if the
measured specific energy loss $\mathrm{d}E/\mathrm{d}x$ of the track
is consistent with the expectation value for the kaon mass hypothesis,
i.e., if the corresponding confidence level $P_{\mathrm{K}}$ is greater than 10\%.
 The track is classified as a pion if $P_{\pi}$ is at least 1$\%$.
 Thus, each track can be flagged as a kaon or pion or both or neither.
 
The $\pi^0$ candidates are formed from pairs of photons found in ECAL
 with an energy of at least $250\unit{MeV}$ each and an 
 invariant mass within $85\unit{MeV}/c^{2}$ of the nominal $\pi^0$ mass. 
 In order to improve the energy resolution of these
$\pi^0$'s the energies of the photons are refitted using the $\pi^0$ mass
as constraint. If the confidence level of this fit is greater than 5\% and 
if $|\cos\theta_{\pi^0}| < 0.93$, where $\theta_{\pi^0}$ is the polar angle 
of the $\pi^0$ candidate with respect to the beam axis, the $\pi^0$ candidate 
is retained.

\begin{figure}[btp]
  \begin{center}
    \includegraphics[width=.8\textwidth]{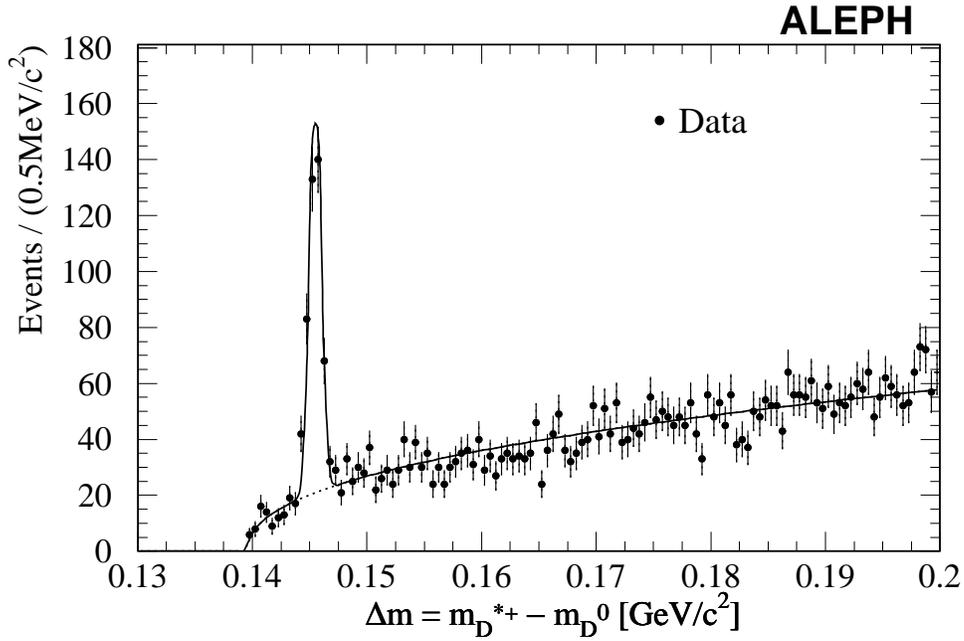}
    \caption{Mass difference of reconstructed D$^{*+}$ and D$^0$
      candidates for all considered D$^0$ decay modes together. 
      The points show data, the error bars represent statistical 
      uncertainties, and the solid curve indicates the result of 
      an unbinned maximum likelihood fit.}
    \label{fig:dm_tot}
  \end{center}
\end{figure}

\begin{figure}[btp]
  \begin{center}
    \includegraphics[width=.8\textwidth]{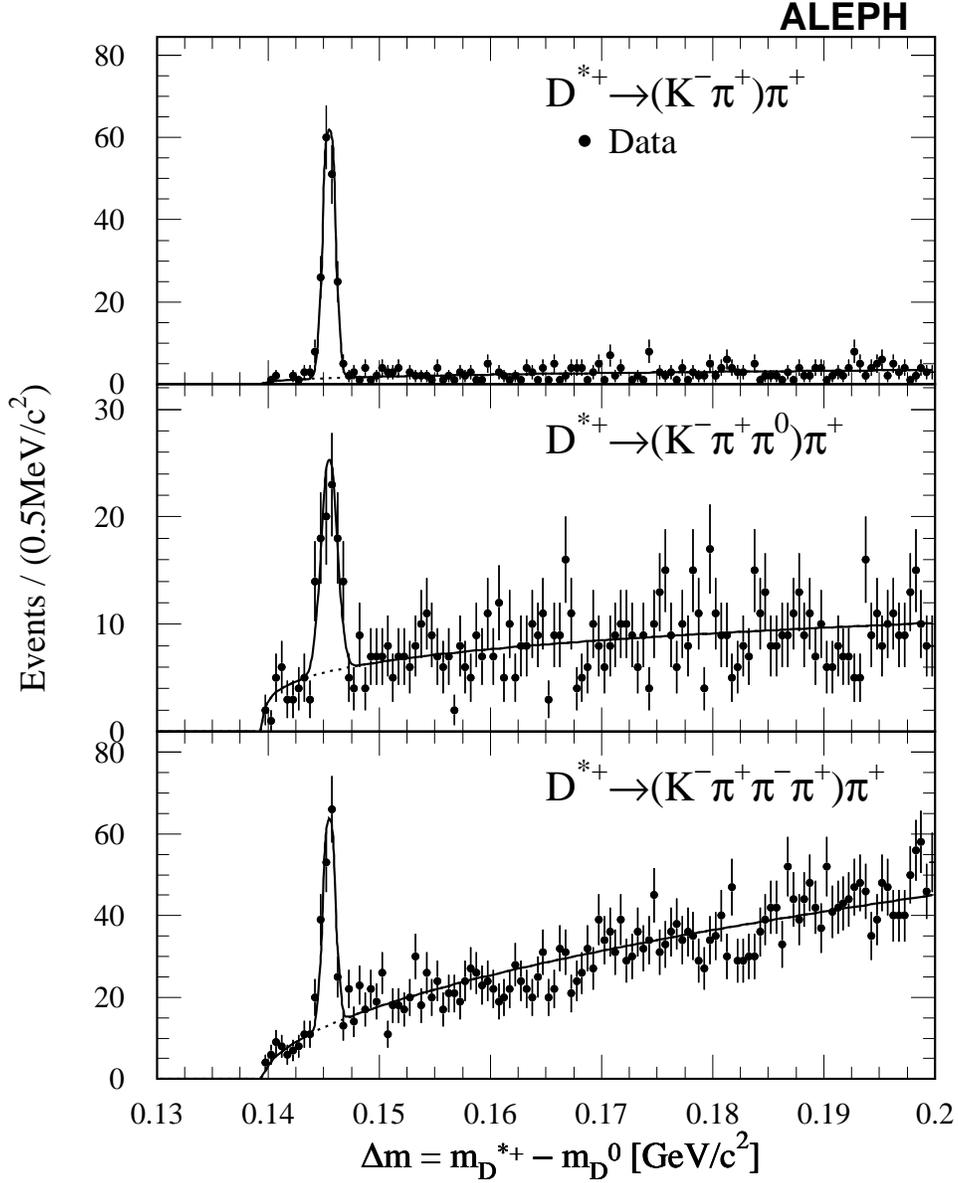}
    \caption{Mass difference of reconstructed D$^{*+}$ and D$^0$
      candidates for three considered D$^0$ decay modes separately. 
      The points show data, the error bars represent statistical 
      uncertainties, and the solid curves indicate the result of 
      unbinned likelihood fits.}
    \label{fig:dm_3modes}
  \end{center}
\end{figure}

The D$^0$ candidates are formed from appropriate combinations of
identified kaons and pions according to three considered decay modes.
 The D$^0$ candidate is retained if it has an invariant mass within 
 20$\unit{MeV}/c^2$, 65$\unit{MeV}/c^2$, and 20$\unit{MeV}/c^2$ 
 of the nominal D$^0$ mass for decay mode (1), (2), and (3),
  respectively. These mass ranges correspond to about three times the mass
   resolution.
 In order to reduce the combinatorial background in mode (3), 
 the four tracks composing the D$^0$ are fitted to a common vertex 
 and the confidence level of this fit is required to be greater than 0.2\%.
The combination of each D$^0$ with one of the remaining $\pi^+$ candidates is considered to be a
D$^{*+}$ candidate. In order to reduce combinatorial background from soft
processes and to limit the kinematic range of the D$^{*+}$ to
the acceptance range of the detector with reasonable efficiency, cuts
were applied to the transverse momentum $p_{\rm t}$ and the pseudorapidity
$\eta = -\ln (\tan (\theta /2))$ of the D$^{*+}$:
\begin{equation}
  \label{eq:acceptance_range}
  2\unit{GeV}/c < p_{\rm t}^{\rm D^{*+}} < 12\unit{GeV}/c\,,\quad |\ett| < 1.5 \quad .
\end{equation}
If there are several D$^{*+}$ candidates found in one event the
corresponding D$^0$ candidates are compared in mass and only the candidate
 with D$^0$ mass nearest the nominal D$^0$ mass is retained. If two or more D$^{*+}$ 
 candidates share the same D$^0$ candidate, all of them are retained.
Figure~\ref{fig:dm_tot} shows the mass difference $\Delta m = m_{\rm D^{*+}} - m_{\rm D^0}$
for the selected D$^{*+}$ candidates for all three decay modes together. 
The spectrum rises at the lower threshold given by the pion mass. 
A clear peak is seen around $145.5\unit{MeV}/c^{2}$. In order to extract the number of
D$^{*+}$ events the data distribution is fitted with the following
parametrization:
\begin{equation}
  \label{eq:dm_fitfunc}
  F(\Delta m)  =
    N  \left[
    \frac{1}{\sqrt{2\pi}\sigma}
    {\rm exp} \bigg \{ - \frac{1}{2}
      \left(
        \frac{\Delta m - 145.5\unit{MeV}/c^{2}}{\sigma}
      \right)^2 \bigg \}
    + C
    \left (
      \Delta m - m_{\pi^+}
    \right )^P
    \right ] \quad .
\end{equation}
In order to exclude systematic binning effects an unbinned maximum likelihood 
fit is performed where $C$ and $P$ are used as free parameters.
The normalization $N$ follows from the constraint that the integral of
$F(\Delta m)$ over the range of the fit, 
$ 130 \unit{MeV}/c^{2} < \Delta m < 200 \unit{MeV}/c^{2} $, must be equal to 
the number of entries in the histogram. The width $\sigma$ of the Gaussian
describing the peak is fixed to $0.5\unit{MeV}/c^{2}$, as determined in 
Monte Carlo. The number of D$^{*+}$ 
events is then obtained by integrating the Gaussian part of 
(\ref{eq:dm_fitfunc}) in the range of $145.5 \unit{MeV}/c^{2} \pm 3\sigma$.
%In the data sample analyzed
As the result a total of $ 360.0 \pm 27.0_{\mathrm{stat}} $ D$^{*+}$ events are observed 
for all three D$^{*+}$
decay modes together. Among the possible background processes,
 only the contribution from  
 $\gamma\gamma \rightarrow \bbbar \rightarrow \rm D^{*\pm} X $
is found to be sizeable. This contribution is estimated to be 
$ 20.5 \pm 1.6_{\mathrm{stat}} $ D$^{*+}$ events from a  
 $\gamma\gamma \rightarrow \bbbar \rightarrow \rm D^{*\pm} X $ Monte Carlo 
 sample and the total cross section $\sigma ( \epem \rightarrow \epem \bbbar )$
 measured in \cite{l3:bb}. 
 After subtraction of this background, a total of $ 339.5 \pm 27.0_{\mathrm{stat}} $ 
 D$^{*+}$ events are found in the data sample analyzed. The mass difference distributions 
 for three channels separately are shown in Fig.~\ref{fig:dm_3modes}.
\section{Cross Section Measurements}
\label{sec:XS}
\subsection{Relative Fractions of Direct and Single-resolved Contributions}
\label{sec:separation}
As mentioned in the introduction, open charm production in
$\gamma\gamma$ collisions is dominated by contributions from
direct and single-resolved processes.
 In the direct case the $\ccbar$ pair makes up the final state 
 of the $\gamma\gamma$ system (in LO) whereas in the single-resolved case 
 the partons of the resolved photon (photon residue) in addition to 
 the $\ccbar$ pair make up the final state. The transverse momentum
 $p_{\rm t}^{\rm D^{*+}}$ of the D$^{*+}$ is correlated with
  the invariant mass of the $\ccbar$ system and the total visible
  invariant mass $W_{\mathrm{vis}}$ is in turn correlated with the
  invariant mass of the total $\gamma\gamma$ system. The ratio
    $p_{\rm t}^{\rm D^{*+}} /W_{\mathrm{vis}}$
  should therefore  be distributed at higher values for the direct case compared
  to the distribution of single-resolved events.

\begin{figure}[htbp]
  \begin{center}
    \includegraphics[width=0.68\textwidth]{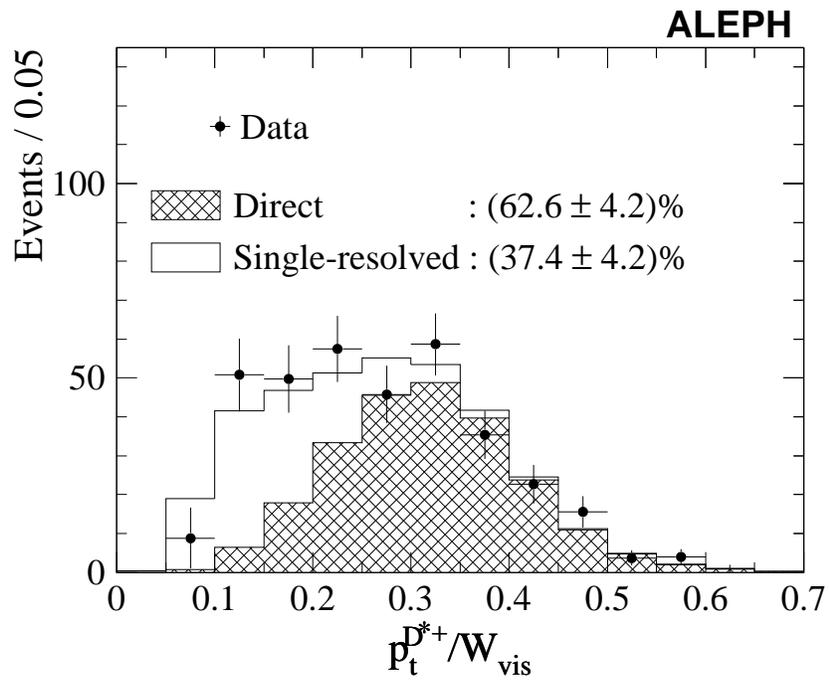}
    \caption{$p_{\rm t}^{\rm D^{*+}}$/$W_{\textrm{vis}}$ distribution for reconstructed
      D$^{*+}$ events. The points with error bars show data. The relative contributions from direct (shaded  histogram)
            and single-resolved (open histogram) processes are extracted by
            means of a fit.}
    \label{fig:sepa_ptwvis}
  \end{center}
\end{figure}

Figure~\ref{fig:sepa_ptwvis} shows the distribution of $ p_{\rm t}^{\rm D^{*+}} /
W_{\mathrm{vis}} $ in data for all events found in the signal region
of the mass-difference spectrum. Combinatorial background has been 
subtracted using events of the upper sideband 
$ 0.16\unit{GeV}/c^{2} < \Delta m < 0.2\unit{GeV}/c^{2} $
of the mass-difference spectrum. Background from $\bbbar$ production
has also been subtracted. The relative fractions are determined by
fitting the sum of the direct and single-resolved Monte Carlo distributions 
to data with the relative fraction as a free parameter of the fit.
The total number of entries in this Monte Carlo sum is required to be
equal to the number of entries in the data distribution.
The fit yields a direct contribution of
$r_{\mathrm{dir}} = ( 62.6 \pm 4.2 )\% $ and a single-resolved
contribution of $r_{\mathrm{res}} = 1 - r_{\mathrm{dir}} = ( 37.4 \pm 
4.2 ) \% $.

\subsection{Differential Cross Sections}
\label{sec:diff_XS}
Two differential cross sections for the production of D$^{*+}$ mesons are
determined: the first one as a function of the transverse D$^{*+}$ 
momentum $p_{\rm t}^{\rm D^{*+}}$, and the second as a function of  
pseudorapidity $|\ett|$. Both are restricted to the range defined in 
Eq.~(\ref{eq:acceptance_range}).
 The former is measured in three $p_{\rm t}^{\rm D^{*+}}$ bins: 
 [2--3], [3--5], [5--12] $\unit{GeV}/c$, and the latter in three $|\ett|$ bins:
[0--0.5], [0.5--1.0], [1.0--1.5]. All considered D$^{*+}$ decay modes were 
treated separately. 

The average differential cross section \dsdptt\ for a given 
$p_{\rm t}^{\rm D^{*+}}$ bin and $|\ett| < 1.5$ is obtained by

\begin{equation}
\label{eq:diff_xs}
  \frac{\mathrm{d}\sigma}{\mathrm{d}p_{\mathrm{t}}^{\mathrm{D}^{\ast +}}
    }  = 
  \frac{
    N^{\mathrm{D}^{*+}}_{\mathrm{found}} }{
  \Delta p_{\rm t}^{\rm D^{*+}} 
  \mathcal{L}
  B_* 
  B_0 
  \epsilon_{\ptt}}\qquad .
\end{equation}
Analogously one obtains \dsdetat\ for a given bin in $ |\ett| $ and 
$2\unit{GeV}/c < p_{\rm t}^{\rm D^{*+}} < 12\unit{GeV}/c$
\begin{displaymath}
  \frac{\mathrm{d}\sigma}{\mathrm{d}|\eta^{\mathrm{D}^{\ast +}}|
    }  = 
  \frac{
    N^{\mathrm{D}^{*+}}_{\mathrm{found}} }{
  \Delta |\ett| 
  \mathcal{L} 
  B_* 
  B_0 
  \epsilon_{|\ett|}}\qquad ,
\end{displaymath}
where
\begin{itemize}
\itemsep -4pt
\item $N^{\mathrm{D}^{*+}}_{\mathrm{found}}$ is the number of 
  $\mathrm{D}^{*+}$ found in the considered bin after subtracting the $\bbbar$ 
  background (determined as described in Section~\ref{sec:dstar}) with the width
  of the fitted Gaussian (\ref{eq:dm_fitfunc}) being fixed to 
  $0.5\unit{MeV}/c^{2}$ for decay modes (1) and (3) and 
  $0.7\unit{MeV}/c^{2}$ for decay mode (2),
\item $\Delta p_{\rm t}^{\rm D^{*+}}$, $\Delta |\ett|$ are the considered intervals in $p_{\rm t}^{\rm D^{*+}}$ and
  $ | \ett | $,
\item $\mathcal{L} = 699\unit{pb^{-1}}$ is the integrated luminosity of the data analyzed,

 \item $B_*$ is the branching ratio BR$(\mathrm{D}^{*+} \rightarrow \mathrm{D}^0
  \pi^+) = ( 68.3 \pm 1.4 ) \% $ \cite{pdg1998},
\item $B_0$ is the branching ratio of the considered D$^0$ decay mode
  \cite{pdg1998},
\item  $\epsilon_{\ptt}$ ($\epsilon_{|\ett|}$) is the efficiency of reconstructing a 
D$^{*+}$ candidate in the given $p_{\rm t}^{\rm D^{*+}}$ ($|\ett|$) bin in the 
considered decay mode.  Since efficiencies are determined separately for direct
  and single-resolved processes ($\epsilon_{\ptt}^{\mathrm{dir}}$ and
  $\epsilon_{\ptt}^{\mathrm{res}}$, respectively) the total efficiency 
 is a weighted combination using the fractions
 as determined in Section~\ref{sec:separation},
 \begin{displaymath}
   \epsilon_{\ptt} = r_{\mathrm{dir}} \epsilon_{\ptt}^{\mathrm{dir}}
             + r_{\mathrm{res}} \epsilon_{\ptt}^{\mathrm{res}} \quad ,  
  \end{displaymath} 
 \begin{displaymath}          
   \epsilon_{|\ett|} = r_{\mathrm{dir}} \epsilon_{|\ett|}^{\mathrm{dir}}
             + r_{\mathrm{res}} \epsilon_{|\ett|}^{\mathrm{res}} \quad .
  \end{displaymath} 
\end{itemize}
Tables~\ref{tab:diff_pt} and \ref{tab:diff_eta} show the number of
 D$^{*+}$ mesons found in the chosen $p_{\rm t}^{\rm D^{*+}}$ and
$|\ett| $ bins respectively, as well as the derived differential cross
sections $\mathrm{d} \sigma / \mathrm{d} p_{\mathrm{t}} ^ {\mathrm{D}
  ^ {\ast +}}$ and $\mathrm{d} \sigma / \mathrm{d} | \eta ^ {\mathrm{D}
  ^ {\ast +}} |$ with their statistical and systematic errors.
    The resulting cross sections for the different D$^{*+}$
decay modes are consistent with each other for all bins in $p_{\rm t}^{\rm D^{*+}}$ as well
as in $ | \ett | $, taking into account the statistical uncertainties.
The weighted average over all of the considered D$^{*+}$ decay modes is
given in Table~\ref{tab:diff_comb} for each $p_{\rm t}^{\rm D^{*+}}$ and $ | \ett| $ bin,
 where only the dominating statistical uncertainties are used for weighting.

\subsubsection{Systematic Errors on Differential Cross Sections}
\label{sec:syst_diff}

The study of systematic errors was performed separately for each $p_{\rm t}^{\rm D^{*+}}$ 
and $| \ett |$ bin and for each of the considered D$^{*+}$ decay modes,
unless otherwise specified.

 The systematic error introduced by the event selection was studied
by varying the cuts within the resolution obtained from the Monte Carlo 
detector simulation. The systematic
uncertainty was estimated from the resulting relative variation of the
efficiency. This yields an uncertainty of 0.6\%--6.4\%,
depending on the considered  $p_{\rm t}^{\rm D^{*+}}$ or $| \ett |$ bin and on the D$^{*+}$
decay mode.

 The selection of pion and kaon candidates depends essentially on the
$\mathrm{d}E / \mathrm{d}x$ measurement as well as on
the expectation values $\mathrm{d}E/\mathrm{d}x_{\rm{exp,h}}$
 used to calculate the probability for a given mass
hypothesis $m_{\rm h}$. The uncertainty of
the $\mathrm{d}E / \mathrm{d}x$ calibration changes the efficiency by 
0.5\%--5.8\%. These deviations are used as an estimate of the systematic error.

 The systematic error due to the accepted mass range used to classify D$^0$ 
 candidates was examined by comparison of the mass distributions of D$^0$
candidates which contributed to the D$^{*+}$ signal in data and Monte
Carlo for each D$^0$ decay mode separately. A Gaussian fit was applied
to these distributions. The fraction of the fitted Gaussian which
lies within the accepted mass range differs between data
and Monte Carlo by less than $ 0.6\% $. Thus, the uncertainty due to this source
 can be neglected.

 In order to estimate the error introduced by the method for extracting
the number of D$^{*+}$ events (Section~\ref{sec:dstar})
the mean of the fitted Gaussian in Eq.~(\ref{eq:dm_fitfunc}) 
was varied by $\pm 0.05\unit{MeV}/c^{2}$, and the width was varied by 10\%
 about its values as obtained in Monte Carlo.
 The resulting relative error on the efficiencies was 0.8\%--2.1\%.

 A variation of the interval that defines the upper sideband yields 
a variation in $r_\mathrm{dir}$ of less than $0.05\%$. 
Hence, this source is negligible. The present analysis assumes the fraction 
$r_\mathrm{dir}$ to be constant over the
 considered kinematic range. Monte Carlo studies show a variation of this 
 fraction of up to 12\% in this range, depending on the bin in $p_{\rm t}^{\rm D^{
*+}}$ and $| \ett |$. 
 A relative uncertainty of 10\% is therefore added in quadrature to 
the statistical uncertainty of $r_\mathrm{res}$. 
A variation of $r_\mathrm{dir/res}$ within these uncertainties
yields a variation in the cross section of 0.3\%--3.4\%, which is used 
to estimate the introduced uncertainty.

The statistical error of $\bbbar$ background subtraction and the uncertainties 
of the total cross section $\sigma ( \epem \rightarrow \epem \bbbar )$ yield a 
systematic error of 1.2\%--3.4\% on the differential cross sections.

The overall trigger efficiency of the selected
D$^{*+}$ events is estimated to be consistent with $100\%$ with a
statistical uncertainty of $1\%$. Thus no correction is made for this source.

 The relative errors on the branching ratios given in \cite{pdg1998} 
 are used to estimate the corresponding relative systematic uncertainties in the cross sections.

 Similarly the relative uncertainties in the efficiencies due to finite
statistics in the Monte Carlo samples, 0.5\%--2.3\%, are taken into account.

All systematic errors are assumed to be uncorrelated and therefore
added in quadrature. Table~\ref{tab:syst_err} shows a summary of the systematic 
uncertainties.
\vspace{-3mm}
\subsubsection{Comparison to Theory}
\label{sec:comp_nlo}

\begin{figure}[htbp]
  \begin{center}
    \includegraphics[width=.8\textwidth]{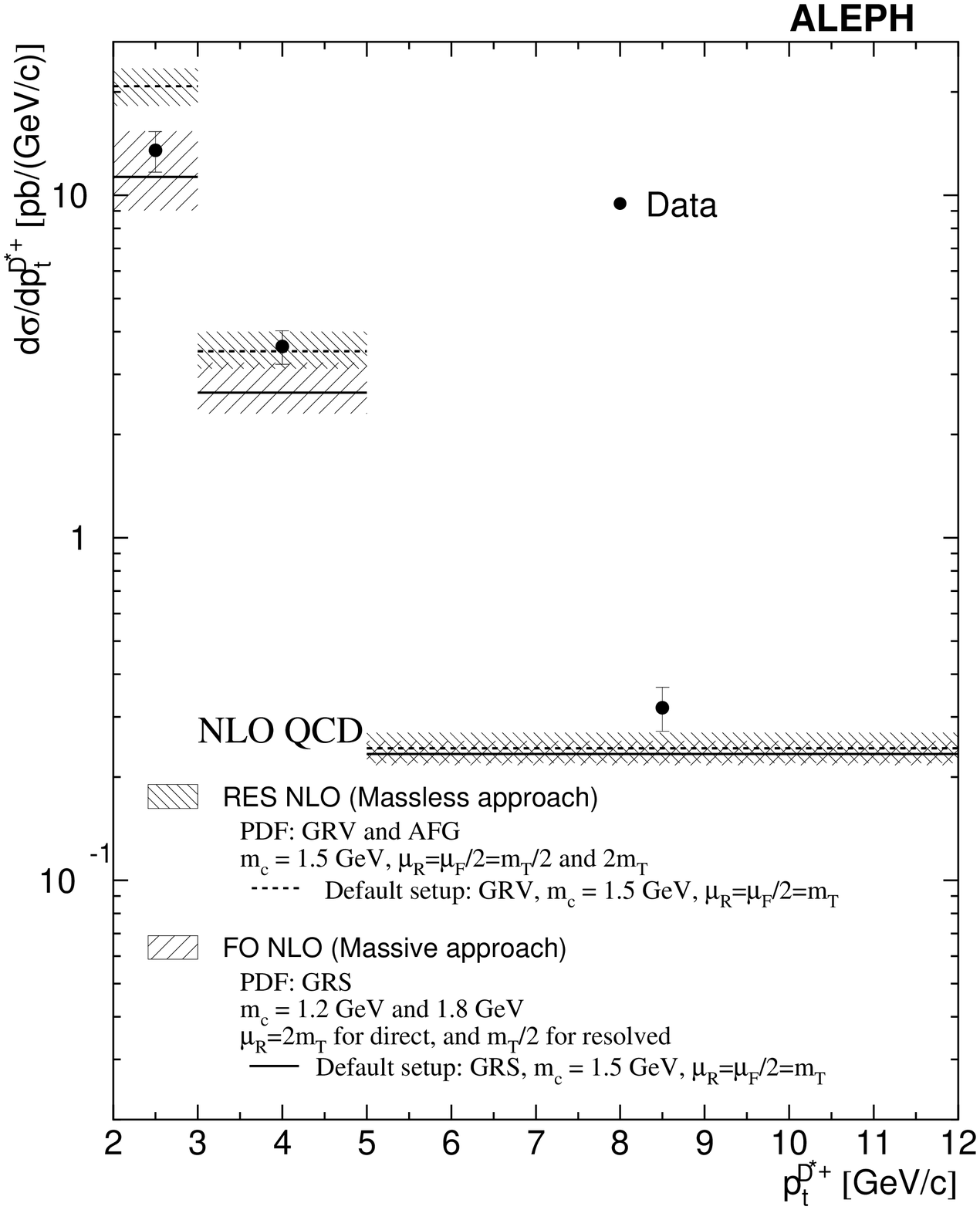}
    \caption{Differential cross section \dsdptt\ for the inclusive D$^{*+}$ 
             production. The points show the combined differential cross 
             sections from the three decay modes under studies.
             The error bars correspond to the quadratic sum of statistical 
             and systematic uncertainties. 
             The data are compared to the fixed-order (FO) NLO 
             \cite{frix:dstar} and the resummed (RES) NLO \cite{kniehl:dstar} 
             calculations shown as the solid and dashed lines, respectively.
              The shaded bands represent the theoretical 
             uncertainties of these calculations.}
    \label{fig:diff_pt}
  \end{center}
\end{figure}

\begin{figure}[htbp]
  \begin{center}
    \includegraphics[width=.8\textwidth]{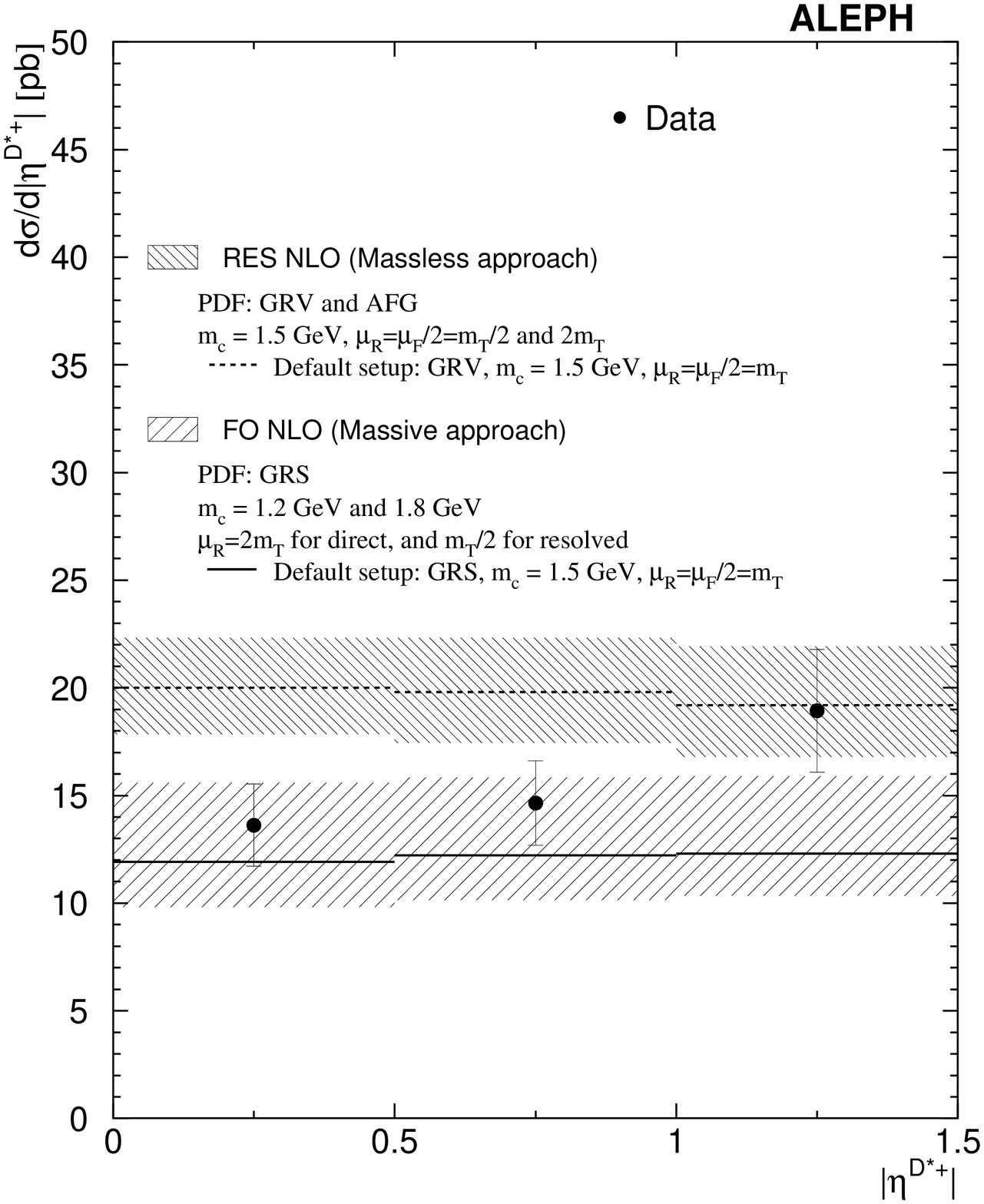}
    \caption{Differential cross section \dsdetat\ for the inclusive D$^{*+}$ 
             production. The points show the combined differential cross 
             sections from the three decay modes under studies.
             The error bars correspond to the quadratic sum of statistical 
             and systematic uncertainties. 
             The data are compared to the fixed-order (FO) NLO 
             \cite{frix:dstar} and the resummed (RES) NLO \cite{kniehl:dstar} 
             calculations shown as the solid and dashed lines,  
             respectively. The shaded bands represent the theoretical 
             uncertainties of these calculations.}
    \label{fig:diff_et}
  \end{center}
\end{figure}

Figures~\ref{fig:diff_pt} and \ref{fig:diff_et} show the measured 
\dsdptt\ and \dsdetat\ in comparison to two different NLO perturbative QCD 
calculations, the fixed-order (FO) NLO (also known as massive approach) 
\cite{frix:dstar} and the resummed (RES) NLO (massless approach) 
\cite{kniehl:dstar}. In both cases, the  charm quark mass
$ m_{\mathrm{c}} $ is set to $ 1.5\unit{GeV/c^2} $, the renormalization scale 
$\mu_{\rm R}$ and the factorization scale $ \mu_{\rm F}$ are chosen such that 
$\mu_{\rm F}^2=4 \mu_{\rm R}^2 = m_{\rm T}^2 \equiv m_{\mathrm{c}}^2 + p_{\rm t}(\mathrm{c})^2 $,
 where  $p_{\rm t}(\mathrm{c})$ is the transverse momentum  of the
 charm quark. For the resolved contribution the photonic parton densities of 
 the GRS-HO parametrization are chosen \cite{GRS-HO} in the FO NLO calculation,
whereas the RES NLO uses GRV-HO \cite{GRV-HO}. The fragmentation of the
charm quark to the D$^{*+}$ is modelled by the fragmentation function
suggested by Peterson et al.~\cite{peterson}, with 
$\epsilon_{\mathrm{c} } = 0.035 $ in the case of FO NLO. The RES NLO calculation 
uses $\epsilon_{\mathrm{c} } = 0.185 $, which was determined by using 
nonperturbative fragmentation functions fitted \cite{kniehl:dstar} to ALEPH 
measurements of inclusive D$^{*+}$ production in $\epem$ annihilation 
\cite{ALEPH:D*_e+e-}. 
The results of the two NLO QCD calculations are represented by the dashed
lines (for RES NLO) and solid lines (for FO NLO) in both 
Fig.~\ref{fig:diff_pt} and Fig.~\ref{fig:diff_et}. In order to estimate the 
theoretical uncertainties, the FO NLO calculation was repeated with the   
charm mass and the renormalization scale varied as described in the figures.
The RES NLO calculation is also repeated using the AFG \cite{AFG} ansatz as an 
alternative for parton density function and 
varying the renormalization and factorization scales.
The resulting theoretical uncertainties are indicated by the bands around the
corresponding default values in Figs.~\ref{fig:diff_pt} and \ref{fig:diff_et}. 

 Altogether, the measurement of \dsdptt\ seems to favour a harder 
$p_{\rm t}^{\rm D^{*+}}$ spectrum than predicted. The RES NLO calculation clearly 
overestimates the measurement in the low $\ptt$ region, while the FO NLO 
calculation slightly underestimates it in the $\ptt > 3.0\unit{GeV}/c$ region. 
 The measured \dsdetat\ is consistent with the almost flat distribution 
predicted by both NLO calculations, but the measurement of \dsdetat\ 
is again overestimated by the RES NLO calculation and somewhat underestimated 
by the FO NLO calculation.
\subsection{Visible Cross Section}
\label{sec:Vis_XS}
The visible cross section $\sigma_{\mathrm{vis}}^{\mathrm{D}^{*+}} 
( \epem  \rightarrow \epem \mathrm{D}^{*+} X )$ is calculated separately in the 
acceptance range [Eq.~(\ref{eq:acceptance_range})] for the three considered 
decay modes by
\begin{equation}
  \label{eq:vis_xs}
  \sigma_{\mathrm{vis}}^{\mathrm{D}^{*+}} ( \epem  \rightarrow \epem
  \mathrm{D}^{*+} X ) =  
  \frac{
    N^{\mathrm{D}^{*+}}_{\mathrm{found}} }{
  \mathcal{L} 
  B_* 
  B_0 
  \epsilon}\qquad, \,
\end{equation}
where the notation is as the same as in Eq.~(\ref{eq:diff_xs}). The numbers of 
$\mathrm{D}^{*+}$ found and the efficiencies of reconstructing a D$^{*+}$
  candidate for direct and single-resolved processes are listed in 
  Table~\ref{tab:sigma_vis} together with the derived visible cross sections
  $\sigma_{\mathrm{vis}}^{\mathrm{D}^{*+}} ( \epem  \rightarrow \epem
  \mathrm{D}^{*+} X )$ and their uncertainties for the three decay modes. 
  The systematic error is determined in the same way as for
   differential cross sections (Section~\ref{sec:syst_diff}). 
   The weighted average over all of the considered decay modes using 
   the dominating statistical uncertainties for weighting is 
\begin{equation}
  \label{eq:sig_integr_meas}
  \sigma_{\mathrm{vis}}^{\mathrm{D}^{*+}} ( \epem  \rightarrow \epem \mathrm{D}^{*+} X )
   = 23.39 \pm 1.64_{\mathrm{stat}} \pm  1.52_{\mathrm{syst}} \, \unit{pb} \quad .
\end{equation}
The theoretically predicted cross section \cite{frix:dstar} is 
\begin{equation}
  \label{eq:sig_integr_theo}
  \sigma_{\mathrm{vis}}^{\mathrm{D}^{*+}} ( \epem  \rightarrow \epem
  \mathrm{D}^{*+} X ) = 17.3
  \raisebox{.15ex}{\footnotesize$\arraycolsep=0pt\def\arraystretch{1}
%  \mbox{\small$\arraycolsep=0pt\def\arraystretch{1}
  \begin{array}{l}
    + 5.1\\-2.9
  \end{array}$} \, \unit{pb}  \quad ,
\end{equation}
and is consistent with this measurement within the given
uncertainties. 
\subsection{Total Cross Section}
\label{sec:Total_XS}
The total cross section 
for the reaction $ \epem \rightarrow \epem \ccbar $ is given by
\begin{equation}
  \label{eq:sigma_tot}
%  \sigma_{\mathrm{tot}}^{\mathrm{c\bar{c}}} =
  \sigma ( \epem \rightarrow \epem \ccbar ) =
  \frac{\sigma_{\mathrm{vis}}^{\mathrm{D}^{*+}}} {2 
    P_{\mathrm{c} \rightarrow \mathrm{D}^{*+}}} 
  ( r_{\mathrm{dir}}  R_{\mathrm{dir}} +
    r_{\mathrm{res}}  R_{\mathrm{res}}  ) \quad ,
\end{equation}
where the symbols are as follows:
\begin{itemize}
\itemsep -4pt
\item $ \sigma_{\mathrm{vis}}^{\mathrm{D}^{*+}} $ is the
  visible inclusive D$^{*+}$ cross section determined in the
  previous section;
\item $ P_{\mathrm{c} \rightarrow \mathrm{D}^{*+}} $ is the
  probability for a charm quark to fragment into a D$^{*+}$ meson (taking
  the combined quantity $ P_{\mathrm{c} \rightarrow
    \mathrm{D}^{*+}} \times \rm BR(\mathrm{D}^{*+} \rightarrow
  \mathrm{D}^0 \pi^+) = 0.1631 \pm 0.0050$ from \cite{LEP_EW} and
  using $\rm BR(\mathrm{D}^{*+}
  \rightarrow \mathrm{D}^0 \pi^+) = ( 68.3 \pm 1.4 ) \% $ 
  \cite{pdg1998} yields $ P_{\mathrm{c} \rightarrow
    \mathrm{D}^{*+}} = 0.2388 \pm 0.0088 $);
 
 \item the factor 2 in the denominator takes into account that, for
  the single inclusive cross sections, both the D$^{*+}$ and the
  D$^{*-}$ mesons were counted;
\item $ r_{\mathrm{dir}} $ and $ r_{\mathrm{res}} $ are the fractions
  of the direct and single-resolved contributions in the considered
  acceptance range, as described in Section~\ref{sec:separation};
\item $ R_{\mathrm{dir}} $ is the ratio
  \begin{displaymath}
    R_{\mathrm{dir}} =
    \frac{\sigma_{\mathrm{tot,dir}}^{\mathrm{D}^{*+}}}
         {\sigma_{\mathrm{vis,dir}}^{\mathrm{D}^{*+}}}
  \end{displaymath}
  of the total D$^{*+}$ cross section to the visible cross section in
  the range of Eq.~(\ref{eq:acceptance_range}) for direct processes.
  It describes the extrapolation of the measured cross section to the
  total phase space available. $ R_{\mathrm{res}} $ is the
  corresponding quantity for the single-resolved case.
\end{itemize}
Separate Monte Carlo samples are used to estimate $ R_{\mathrm{dir}} $ and
 $ R_{\mathrm{res}} $ for direct and single-resolved processes.
The parameters used to determine $ R_{\mathrm{dir}} $ and $
R_{\mathrm{res}} $ are described in Section~\ref{sec:MC}. This yields
$R_{\mathrm{dir}} = 12.74 \pm 0.45_{\mathrm{stat}}$ 
 and $R_{\mathrm{res}} = 18.62 \pm 0.80_{\mathrm{stat}}$.

The main theoretical uncertainties entering the calculation of the
extrapolation factors stem from the uncertainty of the 
charm quark mass. 
A variation of the charm mass to $m_{\mathrm{c}} = 1.3\unit{GeV}$ and
$m_{\mathrm{c}} = 1.7\unit{GeV}$ yields relative errors on
$R_{\mathrm{dir}}$ of $ \pm 10\%$ and on $R_{\mathrm{res}}$ of
+43\% and $-19\%$, respectively.

In the single-resolved case an additional uncertainty enters
$R_{\mathrm{res}}$ by the choice of the parton density functions
describing the resolved photon. Alternatively to the default choice
the GRV-LO parametrization \cite{grv-lo} was used to calculate $R_{
  \mathrm{res} }$. This yields a relative deviation of 12\% and is 
 added in quadrature to the other systematic uncertainties on 
 $R_{\mathrm{res}}$.
The following values are therefore obtained:
\begin{eqnarray}
  \label{eq:Rdir}
  R_{ \mathrm{dir} }  =  12.7 \pm 1.3 \nonumber \\   
  R_{ \mathrm{res} }  =  18.6
  \raisebox{.15ex}{\footnotesize$\arraycolsep=0pt\def\arraystretch{1} 
  \begin{array}{c}
    + 8.3\\
    - 4.2
  \end{array}$}\quad .
\nonumber
\end{eqnarray}

The uncertainties in $ r_{ \mathrm{dir} } $, $\sigma_{ \mathrm{vis} }
^ { \mathrm{D}^{*+} }$, and $ P_{ \mathrm{c} \rightarrow \mathrm{D} ^
  {*+} }$, which are assumed to be uncorrelated, are taken into account 
  in the estimation of the statistical and systematic error on the total 
  cross section by Gaussian error propagation.
This procedure yields a total cross section for the reaction $ \epem
\rightarrow \epem \ccbar$ at $\epem$ centre-of-mass energies 
$\sqrt{s} = (183-209)\unit{GeV} $, corresponding to the luminosity 
weighted average of $197\unit{GeV}$, 
\begin{equation}
  \label{eq:sigma_tot_result}
  \sigma ( \epem \rightarrow \epem \ccbar )_{<\sqrt{s}>=197\unit{GeV}}
  = 731 \pm 74_{\mathrm{stat}} \pm 47_{\mathrm{syst}}
  \raisebox{.15ex}{\footnotesize$\arraycolsep=0pt\def\arraystretch{1}   
  \begin{array}{l}
    + 157 \\
    - 86_{\mathrm{extr}}
  \end{array}$}
  \, \unit{pb} \quad .
\end{equation}

Alternatively, the total cross section is determined by means of the NLO 
calculation referenced in the previous section; in this case the
cross section is given by
\begin{equation}
  \label{eq:sigma_tot_frix}
%  \sigma_{\mathrm{tot}}^{\mathrm{c\bar{c}}} =
  \sigma ( \epem \rightarrow \epem \ccbar )  = 
  \frac{\sigma_{\mathrm{vis}}^{\mathrm{D}^{*+}}} 
  {2 P_{\mathrm{c} \rightarrow \mathrm{D}^{*+}}} 
  R_{ \mathrm{tot}} \quad .
\end{equation}
The value $ R _ { \mathrm{tot} } = 22.2$ is extracted from
\cite{frix:dstar} by determining the ratio of the calculated total
charm cross section to the charm cross section calculated for the
visible D$^{*+}$ range considered in the present analysis.  
Variation of the parameters entering the calculation yields
deviations in the range from $ -33\% $ to $ +72\% $, which are used
as an estimate of the systematic error due to the extrapolation. 
This results in a total cross section 
\begin{equation}
  \label{eq:sigma_tot_result_frix}
  \sigma ( \epem \rightarrow \epem \ccbar )_{<\sqrt{s}>=197\unit{GeV}}
  = 1087 \pm 86_{\mathrm{stat}} \pm 70_{\mathrm{syst}}
 \raisebox{.15ex}{\footnotesize$\arraycolsep=0pt\def\arraystretch{1}  
  \begin{array}{l}
    + 783 \\
    - 357_{\mathrm{extr}}
  \end{array}$}
  \,  \unit{pb} \quad .
\end{equation}

\begin{figure}[htbp]
  \begin{center}
    \includegraphics[width=\textwidth]{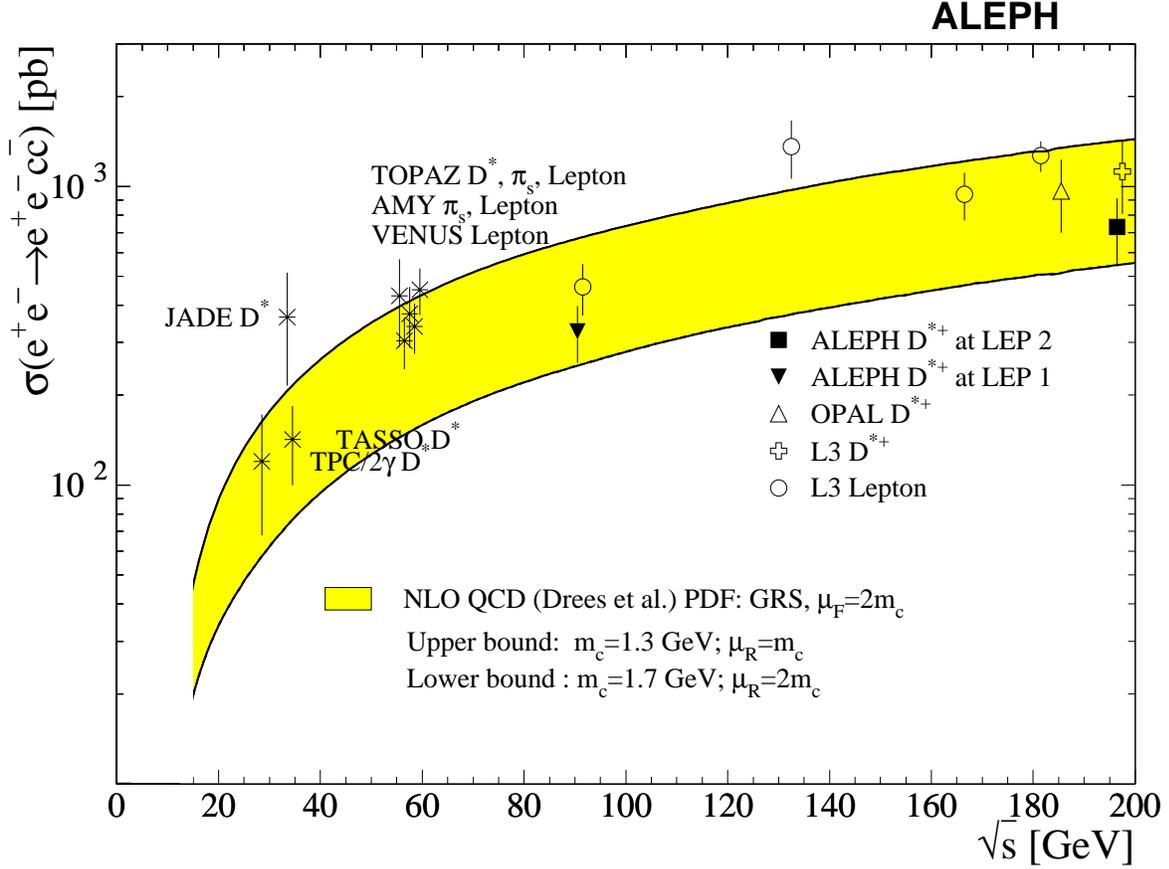}
    \caption{The total cross section for charm production 
     $\sigma ( \epem \rightarrow \epem \ccbar )$ versus the centre-of-mass 
     energy $\sqrt{s}$ of the $ \epem $ system. The measurement of this 
     analysis is shown as a square. The band represents the NLO QCD 
     calculation \cite{zerwas:93}.
     The results obtained by L3 and OPAL using D$^{*+}$ are
       represented in \cite{l3:D*} and \cite{opal:dstar}, respectively.
      The L3 measurements using lepton tag can be found in \cite{l3:lepton}.
      The values for TASSO, TPC/2$\gamma$, JADE, AMY, and VENUS are taken from
      \cite{others:ccbar}.}
    \label{fig:total_sigma}
  \end{center}
\end{figure}

The measured total cross section [Eq.~(\ref{eq:sigma_tot_result})] is shown in 
Fig.~\ref{fig:total_sigma} in comparison to the NLO QCD prediction of Drees
et al.\ \cite{zerwas:93} and to results from other experiments 
\cite{opal:dstar,l3:D*,l3:lepton,others:ccbar}. 
Within the uncertainties, this NLO QCD prediction is in good agreement 
with our measurement and others \cite{armin:01}.

\section{Conclusions}
\label{sec:conclusions}
The inclusive production of D$^{*+}$ mesons in two-photon collisions
was measured using the ALEPH detector at LEP 2 energies in the
reaction D$^{*+} \rightarrow \mathrm{D}^0 \pi^+ $. The D$^0$ mesons
were identified in the decay modes K$^- \pi^+$, K$^- \pi^+
\pi^0$, and K$^- \pi^+ \pi^- \pi^+$. A total of $339.5 \pm 27.0$
D$^{*+}$ events from $\gamma\gamma \rightarrow \ccbar$ was 
found in the kinematic region 
$ 2\unit{GeV}/c < p_{\rm t}^{\rm D^{*+}} < 12\unit{GeV}/c $ and $ |\ett| < 1.5 $.

The fractions of the main contributing processes, direct and single-resolved, 
were determined using the event variable $p_{\rm t}^{\rm D^{*+}} / W_{
  \mathrm{vis} } $ to be $ r_{\mathrm{dir}} = ( 62.6 \pm 4.2 )\% $ and 
 $ r_{\mathrm{res}} = 1 - r_{\mathrm{dir}} = ( 37.4 \pm 4.2 )\% $, 
  within the acceptance. 

The differential cross sections \dsdptt\ and \dsdetat\ were
measured and compared to the fixed-order (FO) NLO
QCD calculation \cite{frix:dstar} and the resummed (RES)
NLO QCD calculation \cite{kniehl:dstar}. 
While the data show a slightly harder spectrum in the $p_{\rm t}^{\rm D^{*+}}$ 
distribution compared to both calculations,
the almost flat distribution of \dsdetat\ which is predicted by the NLO 
calculations for the visible D$^{*+}$ region is in agreement with the 
measurement. Overall, the measurements of \dsdptt\ and \dsdetat\ were slightly 
underestimated by the FO NLO calculation and overestimated by the RES NLO 
calculation.

For the integrated visible D$^{*+}$ cross section a value of $ \sigma ^
{\mathrm{D} ^ {*+} } _ {\mathrm{vis}} = 23.39 \pm 1.64_{\mathrm{stat}} \pm 
1.52_{\mathrm{syst}} \, \unit{pb} $ is obtained which
 is consistent with the FO NLO calculation.

The extrapolation of the visible D$^{*+}$ cross section to the total
cross section of charm production introduces large theoretical
uncertainties and has a large relative uncertainty. 
 Using the LO calculation of the Pythia Monte Carlo we obtain
 \begin{center}
$ \sigma ( \epem \rightarrow \epem \ccbar ) _ {{<\sqrt{s}>=197\unit{GeV}}} =
731 \pm 74_{\mathrm{stat}} \pm 47_{\mathrm{syst}}\,
 \raisebox{.15ex}{\footnotesize$\arraycolsep=0pt\def\arraystretch{1}    
  \begin{array}{l}
    + 157 \\
    - 86_{\mathrm{extr}}
  \end{array}$}
  \, \unit{pb} \quad .$
 \end{center}
A different method using the results from the FO NLO calculation 
\cite{frix:dstar} yields a higher cross section and a larger error.

\subsection*{Acknowledgements}

We wish to thank our colleagues in the CERN accelerator divisions for 
the successful operation of LEP. We are indebted to the engineers and 
technicians in all our institutions for their contribution to the 
excellent performance of ALEPH. Those of us from non-member 
countries thank CERN for its hospitality. We would like to thank 
Stefano Frixione and Bernd Kniehl for fruitful discussions.

\clearpage
\begin{table}[tbp]
  \begin{center}
  \caption{Considered background processes and associated Monte Carlo generators}
\vspace{4pt} 
   \begin{tabular}{l|l}
     
       {Process} & {Monte Carlo Generator} \\  \hline
       $\epem \rightarrow \qqbar$ & PYTHIA 5.7 \cite{sjoestrand:94}\\
       $\epem \rightarrow \tau^+\tau^-$ & KORALZ 4.2  \cite{Jadach:1994}\\
       $\epem \rightarrow \epem\tau^+\tau^-$ & PHOT02 \cite{Vermaseren:1983}\\
       $\epem \rightarrow \rm {W^+W^-}$ &  KORALW 1.21 \cite{Skrzypek:1996}\\ 
     \end{tabular}
	
    \label{tab:backgr_MC}
 \end{center}
\end{table}
\begin{table}[tbp]
  \begin{center}
      \caption{The numbers of D$^{*+}$ mesons found with $|\ett| < 1.5$ in bins 
             of $p_{\rm t}^{\rm D^{*+}}$ for the three decay modes after background subtraction.
              The efficiency is listed separately for direct and single-resolved processes. 
              The differential cross section in bins of the transverse 
              momentum $p_{\rm t}^{\rm D^{*+}}$ of the D$^{*+}$ for each considered D$^{*+}$ decay mode
              is given together with statistical and systematic errors.}
       \vspace{6pt}
     \begin{tabular}{|r@{--}l|r@{$\,\pm\,$}c@{$\,\pm\,$}l|r@{$\,\pm\,$}c@{$\,\pm\,$}l|r@{$\,\pm\,$}c@{$\,\pm\,$}l|}
      \cline{1-11}
      \multicolumn{2}{|c|}{$p_{\rm t}^{\rm D^{*+}}$ range} & 
      \multicolumn{9}{c|}
      {$N_{\mathrm{found}}^{\rm D^{*+}}$} \\ \cline{3-11}
      \multicolumn{2}{|c|}{[GeV/$c$]} &
      \multicolumn{3}{c|}
      {\small$\rm D^{*+} \rightarrow (K^- \pi^+) \pi^+ $} &
      \multicolumn{3}{c|}     
      {\small$\rm D^{*+} \rightarrow (K^- \pi^+ \pi^0) \pi^+$} &
      \multicolumn{3}{c|}      
      {\small$\rm D^{*+} \rightarrow (K^- \pi^+ \pi^- \pi^+) \pi^+$} \\ 
      \hline
      2 & 3  & 69.8 & \multicolumn{2}{@{}l|}{10.7} &  18.7 &
      \multicolumn{2}{@{}l|}{6.2} & 54.5 & \multicolumn{2}{@{}l|}{10.3} \\
      3 & 5  & 72.2 & \multicolumn{2}{@{}l|}{8.1} & 29.0 &
      \multicolumn{2}{@{}l|}{7.8} & 44.9 & \multicolumn{2}{@{}l|}{9.7} \\ 
      5 & 12 & 15.1 & \multicolumn{2}{@{}l|}{3.0} &  20.9 &
      \multicolumn{2}{@{}l|}{5.7} & 29.2 & \multicolumn{2}{@{}l|}{6.8} \\

      \hline 
      \multicolumn{2}{c}{} &
      \multicolumn{3}{c}{} &
      \multicolumn{3}{c}{} &
      \multicolumn{3}{c}{} \\
      \cline{3-11}
      \multicolumn{2}{c|}{} & 
      \multicolumn{9}{c|}
      {Efficiency for direct process $\epsilon_{\ptt}^{\mathrm{dir}}(\%)$} \\ \cline{3-11}
      \multicolumn{2}{c|}{} &
      \multicolumn{3}{c|}
      {\small$\rm D^{*+} \rightarrow (K^- \pi^+) \pi^+ $} &
      \multicolumn{3}{c|}     
      {\small$\rm D^{*+} \rightarrow (K^- \pi^+ \pi^0) \pi^+$} &
      \multicolumn{3}{c|}      
      {\small$\rm D^{*+} \rightarrow (K^- \pi^+ \pi^- \pi^+) \pi^+$} \\ 
      \hline
      2 & 3  & 27.96 & \multicolumn{2}{@{}l|}{0.13} &  2.27 &
      \multicolumn{2}{@{}l|}{0.04} & 11.66 & \multicolumn{2}{@{}l|}{0.09} \\
      3 & 5  & 46.94 & \multicolumn{2}{@{}l|}{0.20} & 6.83 &
      \multicolumn{2}{@{}l|}{0.10} & 24.16 & \multicolumn{2}{@{}l|}{0.17} \\ 
      5 & 12 & 48.73 & \multicolumn{2}{@{}l|}{0.34} &  12.32 &
      \multicolumn{2}{@{}l|}{0.23} & 30.13 & \multicolumn{2}{@{}l|}{0.33} \\ 
    
      \hline 
      \multicolumn{2}{c}{} &
      \multicolumn{3}{c}{} &
      \multicolumn{3}{c}{} &
      \multicolumn{3}{c}{} \\
      \cline{3-11}
      \multicolumn{2}{c|}{} & 
      \multicolumn{9}{c|}
      {Efficiency for single-resolved process $\epsilon_{\ptt}^{\mathrm{res}}(\%)$} \\ \cline{3-11}
      \multicolumn{2}{c|}{} &
      \multicolumn{3}{c|}
      {\small$\rm D^{*+} \rightarrow (K^- \pi^+) \pi^+ $} &
      \multicolumn{3}{c|}     
      {\small$\rm D^{*+} \rightarrow (K^- \pi^+ \pi^0) \pi^+$} &
      \multicolumn{3}{c|}      
      {\small$\rm D^{*+} \rightarrow (K^- \pi^+ \pi^- \pi^+) \pi^+$} \\ 
      \hline
      2 & 3  & 26.81 & \multicolumn{2}{@{}l|}{0.12} &  2.12 &
      \multicolumn{2}{@{}l|}{0.04} & 10.49 & \multicolumn{2}{@{}l|}{0.09} \\
      3 & 5  & 41.95 & \multicolumn{2}{@{}l|}{0.21} & 6.17 &
      \multicolumn{2}{@{}l|}{0.10} & 20.78 & \multicolumn{2}{@{}l|}{0.18} \\ 
      5 & 12 & 34.59 & \multicolumn{2}{@{}l|}{0.41} &  8.8 &
      \multicolumn{2}{@{}l|}{0.24} & 19.83 & \multicolumn{2}{@{}l|}{0.36} \\

      \hline 
      \multicolumn{2}{c}{} &
      \multicolumn{3}{c}{} &
      \multicolumn{3}{c}{} &
      \multicolumn{3}{c}{} \\
      \cline{3-11}
      \multicolumn{2}{c|}{} &
      \multicolumn{9}{c|}
      {$\mathrm{d}\sigma/\mathrm{d}p_{\mathrm{t}}^{\mathrm{D}^{\ast +}}(\unit{pb/GeV}/c)$} \\ \cline{3-11}
      \multicolumn{2}{c|}{} &
      \multicolumn{3}{c|}
      {\small$\rm D^{*+} \rightarrow (K^- \pi^+) \pi^+ $} &
      \multicolumn{3}{c|}     
      {\small$\rm D^{*+} \rightarrow (K^- \pi^+ \pi^0) \pi^+$} &
      \multicolumn{3}{c|}      
      {\small$\rm D^{*+} \rightarrow (K^- \pi^+ \pi^- \pi^+) \pi^+$} \\ 
      \hline

      2 & 3  & 13.80 & 2.12 & 1.04 & 12.70 & 4.21 & 1.20 & 13.38 & 2.51 & 0.89\\
      3 & 5  &  4.36 & 0.49 & 0.22 & 3.32 & 0.90 & 0.27 &  2.70 & 0.58 & 0.17\\
      5 & 12 &  0.27 & 0.05 & 0.01 & 0.41 & 0.11 & 0.03 &  0.44 & 0.10 & 0.03\\
      \hline
    \end{tabular}
    \label{tab:diff_pt}
  \end{center}
\end{table}

\begin{table}[tbp]
  \begin{center}
    \caption{The numbers of D$^{*+}$ mesons found with $ 2\unit{GeV}/c < p_{\rm t}^{\rm D^{*+}} < 12\unit{GeV}/c $ in bins 
             of $|\ett|$ for the three decay modes after background subtraction.
              The efficiency is listed separately for direct and single-resolved processes. 
              The differential cross section in bins of the pseudorapidity $|\ett|$ 
               of the D$^{*+}$ for each considered D$^{*+}$ decay mode
              is given together with statistical and systematic errors.}  
         \vspace{7pt}
     \begin{tabular}{|r@{--}l|r@{$\,\pm\,$}c@{$\,\pm\,$}l|r@{$\,\pm\,$}c@{$\,\pm\,$}l|r@{$\,\pm\,$}c@{$\,\pm\,$}l|}
      \cline{1-11}
      \multicolumn{2}{|c|}{$|\ett|$ range} & 
      \multicolumn{9}{c|}
      {$N_{\mathrm{found}}^{\rm D^{*+}}$} \\ \cline{3-11}
      \multicolumn{2}{|c|}{} &
      \multicolumn{3}{c|}
      {\small$\rm D^{*+} \rightarrow (K^- \pi^+) \pi^+ $} &
      \multicolumn{3}{c|}     
      {\small$\rm D^{*+} \rightarrow (K^- \pi^+ \pi^0) \pi^+$} &
      \multicolumn{3}{c|}      
      {\small$\rm D^{*+} \rightarrow (K^- \pi^+ \pi^- \pi^+) \pi^+$} \\ 
      \hline
        0.0 & 0.5 & 49.2 & \multicolumn{2}{@{}l|}{8.9} &  21.8 &
      \multicolumn{2}{@{}l|}{6.8} & 51.1 & \multicolumn{2}{@{}l|}{10.0} \\
        0.5 & 1.0 & 50.8 & \multicolumn{2}{@{}l|}{8.3} & 26.4 &
      \multicolumn{2}{@{}l|}{7.6} & 45.8 & \multicolumn{2}{@{}l|}{9.5} \\ 
        1.0 & 1.5 & 56.4 & \multicolumn{2}{@{}l|}{7.9} &  18.5 &
      \multicolumn{2}{@{}l|}{6.3} & 29.3 & \multicolumn{2}{@{}l|}{7.6} \\ 
    
      \hline 
      \multicolumn{2}{c}{} &
      \multicolumn{3}{c}{} &
      \multicolumn{3}{c}{} &
      \multicolumn{3}{c}{} \\
      \cline{3-11}
      \multicolumn{2}{c|}{} & 
      \multicolumn{9}{c|}
      {Efficiency for direct process $\epsilon_{|\ett|}^{\mathrm{dir}}(\%)$} \\ \cline{3-11}
      \multicolumn{2}{c|}{} &
      \multicolumn{3}{c|}
      {\small$\rm D^{*+} \rightarrow (K^- \pi^+) \pi^+ $} &
      \multicolumn{3}{c|}     
      {\small$\rm D^{*+} \rightarrow (K^- \pi^+ \pi^0) \pi^+$} &
      \multicolumn{3}{c|}      
      {\small$\rm D^{*+} \rightarrow (K^- \pi^+ \pi^- \pi^+) \pi^+$} \\ 
      \hline
       0.0 & 0.5 & 41.71 & \multicolumn{2}{@{}l|}{0.19} &  5.45 &
      \multicolumn{2}{@{}l|}{0.09} & 20.90 & \multicolumn{2}{@{}l|}{0.16} \\
       0.5 & 1.0 & 39.07 & \multicolumn{2}{@{}l|}{0.19} & 5.24 &
      \multicolumn{2}{@{}l|}{0.08} & 19.70 & \multicolumn{2}{@{}l|}{0.16} \\ 
       1.0 & 1.5 & 27.88 & \multicolumn{2}{@{}l|}{0.17} & 3.72  &
      \multicolumn{2}{@{}l|}{0.07} & 12.19 & \multicolumn{2}{@{}l|}{0.13} \\ 
    
      \hline 
      \multicolumn{2}{c}{} &
      \multicolumn{3}{c}{} &
      \multicolumn{3}{c}{} &
      \multicolumn{3}{c}{} \\
      \cline{3-11}
      \multicolumn{2}{c|}{} & 
      \multicolumn{9}{c|}
      {Efficiency for single-resolved process $\epsilon_{|\ett|}^{\mathrm{res}}(\%)$} \\ \cline{3-11}
      \multicolumn{2}{c|}{} &
      \multicolumn{3}{c|}
      {\small$\rm D^{*+} \rightarrow (K^- \pi^+) \pi^+ $} &
      \multicolumn{3}{c|}     
      {\small$\rm D^{*+} \rightarrow (K^- \pi^+ \pi^0) \pi^+$} &
      \multicolumn{3}{c|}      
      {\small$\rm D^{*+} \rightarrow (K^- \pi^+ \pi^- \pi^+) \pi^+$} \\ 
      \hline
       0.0 & 0.5 & 37.55 & \multicolumn{2}{@{}l|}{0.20} &  4.53 &
      \multicolumn{2}{@{}l|}{0.08} & 17.31 & \multicolumn{2}{@{}l|}{0.16} \\
       0.5 & 1.0 & 34.74 & \multicolumn{2}{@{}l|}{0.19} & 4.16 &
      \multicolumn{2}{@{}l|}{0.08} & 15.95 & \multicolumn{2}{@{}l|}{0.15} \\ 
       1.0 & 1.5 & 24.08 & \multicolumn{2}{@{}l|}{0.16} & 2.75  &
      \multicolumn{2}{@{}l|}{0.06} & 9.66 & \multicolumn{2}{@{}l|}{0.11} \\

      \hline 
      \multicolumn{2}{c}{} &
      \multicolumn{3}{c}{} &
      \multicolumn{3}{c}{} &
      \multicolumn{3}{c}{} \\
      \cline{3-11}
      \multicolumn{2}{c|}{} &
      \multicolumn{9}{c|}
      {$\mathrm{d}\sigma/\mathrm{d}|\eta^{\mathrm{D}^{\ast +}}|\unit{[pb]}$} \\ \cline{3-11}
      \multicolumn{2}{c|}{} &
      \multicolumn{3}{c|}
      {\small$\rm D^{*+} \rightarrow (K^- \pi^+) \pi^+ $} &
      \multicolumn{3}{c|}     
      {\small$\rm D^{*+} \rightarrow (K^- \pi^+ \pi^0) \pi^+$} &
      \multicolumn{3}{c|}      
      {\small$\rm D^{*+} \rightarrow (K^- \pi^+ \pi^- \pi^+) \pi^+$} \\ 
      \hline

     0.0 & 0.5 &  13.33 & 2.40 & 0.85 & 12.86 & 4.02 & 1.10 & 14.40 & 2.80 & 1.00\\
     0.5 & 1.0 &  14.78 & 2.40 & 0.86 & 16.48 & 4.75 & 1.36 & 13.81 & 2.87 & 0.91\\
     1.0 & 1.5 &  23.22 & 3.24 & 2.10 & 16.59 & 5.63 & 1.73 & 14.35 & 3.70 & 1.31\\
      \hline
    \end{tabular}
    \label{tab:diff_eta}
    
  \end{center}
\end{table}

\begin{table}[tbp]
  \begin{center}
     \caption{The combined differential cross sections, 
      $\mathrm{d}\sigma / \mathrm{d}p_{\mathrm{t}} ^
      {\mathrm{D}^{\ast +}}$ and $\mathrm{d}\sigma / \mathrm{d}|
      \eta^{\mathrm{D}^{\ast +}}|$.}
     \vspace{7pt}
     \begin{tabular}{|c|r@{$\,\pm\,$}c@{$\,\pm\,$}l|r@{$\,\pm\,$}c@{$\,\pm\,$}l|r@{$\,\pm\,$}c@{$\,\pm\,$}l|}

      \cline{1-10}
      {} &
      \multicolumn{9}{|c|}
      {$p_{\rm t}^{\rm D^{*+}}$ range $\unit{[GeV}/c]$} \\ \cline{2-10}
      {} &
      \multicolumn{3}{c|}
      {2--3} &
      \multicolumn{3}{c|}     
      {3--5} &
      \multicolumn{3}{c|}      
      {5--12} \\ 
      \hline
    
     $\mathrm{d}\sigma/\mathrm{d}p_{\mathrm{t}}^{\mathrm{D}^{\ast +}}\,\unit{[pb/(GeV}/c)]$ & 13.50  & 1.51 & 1.01 & 3.61 & 0.34 & 0.21 & 0.32 & 0.04 & 0.02\\
       \hline

      {} &
      \multicolumn{3}{c}{} &
      \multicolumn{3}{c}{} &
      \multicolumn{3}{c|}{} \\
      \cline{2-10}
      {} &
      \multicolumn{9}{c|}
      {$|\ett|$ range } \\ \cline{2-10}
      {} &
      \multicolumn{3}{c|}
      {0.0--0.5} &
      \multicolumn{3}{c|}     
      {0.5--1.0} &
      \multicolumn{3}{c|}      
      {1.0--1.5} \\ 
      \hline

     $\mathrm{d}\sigma / \mathrm{d}|\eta^{\mathrm{D}^{\ast +}}|\,[\unit{pb}]$ & 13.62  & 1.65 & 0.94 & 14.65 & 1.71 & 0.94 & 18.93 & 2.23 & 1.75\\
      \hline
    \end{tabular}
     \label{tab:diff_comb}
    
  \end{center}
\end{table}
\begin{table}
  \begin{center}
  \caption{Sources of systematic uncertainty on the differential cross sections.}
 \vspace{7pt}
    \begin{tabular}{l|l}
      Source & Estimated uncertainty \\ \hline\hline
      Event selection & (0.6--6.4)\% \\
      K/$\pi$ selection&  (0.5--5.7)\%  \\
      Accepted mass range for D$^0$  &  $< 0.16\% $, neglected \\
      D$^{*+}$ selection & (0.8--2.1)\%  \\
      D$^{*+}$ from annihilation events & $ < 1\% $, neglected \\
      $\bbbar$ background subtraction & (1.2--3.4)\%  \\
      Fraction of direct/resolved $r_{\mathrm{dir}}$/$r_{\mathrm{res}}$ &(0.3--3.4)\% \\
      BR$(\rm D^{*+} \rightarrow \mathrm{D}^0 \pi^+)$ &  2.0\%\\
      BR$(\rm D^0 \rightarrow \mathrm{K}^- \pi^+)$ &  2.3\%  \\
      BR$(\rm D^0 \rightarrow \mathrm{K}^- \pi^+ \pi^0)$ &  6.5\% \\
      BR$(\rm D^0 \rightarrow \mathrm{K}^- \pi^+ \pi^- \pi^+)$ &  5.3\% \\
      Statistical limitation in Monte Carlo &  (0.5--2.3)\% \\
    \end{tabular}
    \label{tab:syst_err}
  \end{center}
\end{table}

\begin{table}
 \begin{center}
     \caption{The numbers of D$^{*+}$ mesons found in the acceptance range
            $2\unit{GeV}/c < p_{\rm t}^{\rm D^{*+}} < 12\unit{GeV}/c$ and $|\ett| < 1.5$ for 
            the three decay modes after background subtraction.
            The efficiency is listed separately for direct and single-resolved process. 
            The visible cross section $\sigma_{\mathrm{vis}}^{\mathrm{D}^{*+}}$ for each considered D$^{*+}$ decay mode
              is given together with statistical and systematic errors.}
 \vspace{7pt}
     \begin{tabular}{|c|r@{$\,\pm\,$}c@{$\,\pm\,$}l|r@{$\,\pm\,$}c@{$\,\pm\,$}l|r@{$\,\pm\,$}c@{$\,\pm\,$}l|}
       \hline 
    & \multicolumn{3}{c|}
      { $\rm (K^- \pi^+) \pi^+ $} &
      \multicolumn{3}{c|}     
      { $\rm (K^- \pi^+ \pi^0) \pi^+$} &
      \multicolumn{3}{c|}      
      { $\rm (K^- \pi^+ \pi^- \pi^+) \pi^+$} \\ 
      \hline
     {$N_{\mathrm{found}}^{\rm D^{*+}}$} &
       156.4 & \multicolumn{2}{l|}{14.9} &
       67.4 & \multicolumn{2}{l|}{12.3}  &
       128.4 & \multicolumn{2}{l|}{16.3} \\ \hline
     {$\epsilon_{\mathrm{dir}}(\%)$} &
       36.47 & \multicolumn{2}{l|}{0.1} &
       4.81 & \multicolumn{2}{l|}{0.05}  &
       17.71 & \multicolumn{2}{l|}{0.09} \\
     {$\epsilon_{\mathrm{res}}(\%)$} &
       31.68 & \multicolumn{2}{l|}{ 0.1} &
       3.76  & \multicolumn{2}{l|}{0.04}  &
       14.07 & \multicolumn{2}{l|}{0.08} \\ \hline
     {$\sigma_{\mathrm{vis}}^{\mathrm{D}^{*+}}$ (pb)} &
       24.68 & 2.35 & 1.47 &
       23.04 & 4.21 & 1.91 &
       21.76 & 2.76 & 1.41 \\  \hline 
   \end{tabular}
    \label{tab:sigma_vis}
 \end{center}
\end{table}

\clearpage


\end{document}